\documentstyle{mn}

\newcommand{\ltsima}{$\; \buildrel < \over \sim \;$}
\newcommand{\simlt}{\lower.5ex\hbox{\ltsima}}            % < over ~
\newcommand{\gtsima}{$\; \buildrel > \over \sim \;$}
\newcommand{\simgt}{\lower.5ex\hbox{\gtsima}}            % > over ~
\newcommand{\lya}{Ly$\alpha\,$}

\newif\ifAMStwofonts
\ifoldfss
  \ifCUPmtlplainloaded \else
    \NewTextAlphabet{textbfit} {cmbxti10} {}
    \NewTextAlphabet{textbfss} {cmssbx10} {}
    \NewMathAlphabet{mathbfit} {cmbxti10} {} % for math mode
    \NewMathAlphabet{mathbfss} {cmssbx10} {} %  "   "    "
  \fi
  \ifAMStwofonts
    \ifCUPmtlplainloaded \else
      \NewSymbolFont{upmath} {eurm10}
      \NewSymbolFont{AMSa} {msam10}
      \NewMathSymbol{\upi}     {0}{upmath}{19}
      \NewMathSymbol{\umu}     {0}{upmath}{16}
      \NewMathSymbol{\upartial}{0}{upmath}{40}
      \NewMathSymbol{\leqslant}{3}{AMSa}{36}
      \NewMathSymbol{\geqslant}{3}{AMSa}{3E}

       \let\ge=\geqslant
    \fi
  \fi
\fi % End of OFSS

\ifnfssone
  \newmathalphabet{\mathit}
  \addtoversion{normal}{\mathit}{cmr}{m}{it}
  \addtoversion{bold}{\mathit}{cmr}{bx}{it}
  \newmathalphabet{\mathbfit} % math mode version of \textbfit{..}
  \addtoversion{normal}{\mathbfit}{cmr}{bx}{it}
  \addtoversion{bold}{\mathbfit}{cmr}{bx}{it}
  \newmathalphabet{\mathbfss} % math mode version of \textbfss{..}
  \addtoversion{normal}{\mathbfss}{cmss}{bx}{n}
  \addtoversion{bold}{\mathbfss}{cmss}{bx}{n}
  \ifAMStwofonts
    \ifCUPmtlplainloaded \else
      \UseAMStwoboldmath
      \makeatletter
      \new@mathgroup\upmath@group
      \define@mathgroup\mv@normal\upmath@group{eur}{m}{n}
      \define@mathgroup\mv@bold\upmath@group{eur}{b}{n}
      \edef\UPM{\hexnumber\upmath@group}
      \new@mathgroup\amsa@group
      \define@mathgroup\mv@normal\amsa@group{msa}{m}{n}
      \define@mathgroup\mv@bold\amsa@group{msa}{m}{n}
      \edef\AMSa{\hexnumber\amsa@group}
      \makeatother
      \mathchardef\upi="0\UPM19
      \mathchardef\umu="0\UPM16
      \mathchardef\upartial="0\UPM40
      \mathchardef\leqslant="3\AMSa36
      \mathchardef\geqslant="3\AMSa3E

       \let\ge=\geqslant
    \fi
  \fi
\fi % End of NFSS release 1

\ifnfsstwo
  \DeclareMathAlphabet{\mathbfit}{OT1}{cmr}{bx}{it}
  \SetMathAlphabet\mathbfit{bold}{OT1}{cmr}{bx}{it}
  \DeclareMathAlphabet{\mathbfss}{OT1}{cmss}{bx}{n}
  \SetMathAlphabet\mathbfss{bold}{OT1}{cmss}{bx}{n}
  \ifAMStwofonts
    \ifCUPmtlplainloaded \else
      \DeclareSymbolFont{UPM}{U}{eur}{m}{n}
      \SetSymbolFont{UPM}{bold}{U}{eur}{b}{n}
      \DeclareSymbolFont{AMSa}{U}{msa}{m}{n}
      \DeclareMathSymbol{\upi}{0}{UPM}{"19}
      \DeclareMathSymbol{\umu}{0}{UPM}{"16}
      \DeclareMathSymbol{\upartial}{0}{UPM}{"40}
      \DeclareMathSymbol{\leqslant}{3}{AMSa}{"36}
      \DeclareMathSymbol{\geqslant}{3}{AMSa}{"3E}

       \let\ge=\geqslant
    \fi
  \fi
\fi % End of NFSS release 2

\ifCUPmtlplainloaded \else
  \ifAMStwofonts \else % If no AMS fonts
    \def\upi{\pi}
    \def\umu{\mu}
    \def\upartial{\partial}
  \fi
\fi

\title[NICMOS imaging search for damped Ly$\alpha$ galaxies]
{NICMOS imaging search for high-redshift damped Ly$\alpha$ galaxies}
\author[S. J. Warren et al.]
       {S. J. Warren,$^1$ P. M\o ller,$^2$ S. M. Fall, $^3$ P. Jakobsen $^4$\\ 
       $^1$Blackett Laboratory, Imperial College of Science Technology
       and Medicine, Prince Consort Rd, London SW7 2BW \\
       $^2$European Southern Observatory, Karl-Schwarzschild-Strasse 2,
       D-85748 Garching bei M\"{u}nchen, Germany   \\
       $^3$Space Telescope Science Institute, 3700 San Martin Drive,
       Baltimore, MD21218, USA \\
       $^4$Astrophysics Division, European Space Research and
       Technology Centre, 2200 AG Noordwijk, Netherlands}
\date{Accepted
      Received
      in original form}

\begin{document}

\maketitle

\begin{abstract}

We are engaged in a programme of imaging with the STIS and NICMOS
(NIC2) instruments aboard the Hubble Space Telescope (HST), to search
for the galaxy counterparts of 18 high-redshift $z>1.75$ damped \lya
absorption lines and 5 Lyman-limit systems seen in the spectra of 16
target quasars. This paper presents the results of the imaging
campaign with the NIC2 camera. We describe the steps followed in
reducing the data and combining in mosaics, and the methods used for
subtracting the image of the quasar in each field, and for
constructing error frames that include the systematic errors
associated with the psf subtraction. To identify candidate
counterparts, that are either compact or diffuse, we convolved the
image and variance frames with circular top-hat filters of diameter
0.45 and 0.90 arcsec respectively, to create frames of summed $S/N$
within the aperture. For each target quasar we provide catalogues
listing positions and aperture magnitudes of all sources within a
square of side 7.5\arcsec\ centred on the quasar, detected at
$S/N>6$. We find a total of 41 candidates of which three have already
been confirmed spectroscopically as the counterparts. We provide the
aperture magnitude detection limits as a function of impact parameter,
for both detection filters, for each field. The average detection
limit for compact (diffuse) sources is ${\mathrm H_{AB}}=25.0\,
(24.4)$ at an angular separation of $0.56\arcsec\, (0.79\arcsec)$ from
the quasar, improving to ${\mathrm H_{AB}}=25.5\, (24.8)$ at large
angular separations. For the brighter sources we have measured the
half-light radius and the $n$ parameter of the best-fit deconvolved
Sersic-law surface-brightness profile, and the ellipticity and
orientation.

\end{abstract}


\section{Introduction}

The bulk of the neutral gas in the Universe is contained in clouds of
highest column density, ${\mathrm N_{HI}}>2\times 10^{20}$ cm$^{-2}$,
the damped \lya (DLA) absorbers.  The incidence of these objects with
different ${\mathrm N_{HI}}$ in the spectra of background quasars
allows one to determine the comoving density of neutral gas as a
function of redshift or time. At $z\approx$~2--3 this density was
roughly equal to the present comoving density of luminous stars. This
observation has led to the standard interpretation of the DLA
absorbers as the gas reservoirs from which stars form, i.e., as the
interstellar components of galaxies and protogalaxies (Wolfe et
al. 1986). This interpretation is supported by global studies of
chemical evolution, which relate the evolution of the comoving
densities of stars, gas, heavy elements, and dust in galaxies (Pei,
Fall, \& Hauser 1999). This global approach, however, tells us
nothing about the properties of individual DLA galaxies, such as their
luminosities, sizes, and morphologies.

Deep imaging is needed to detect and characterize the stellar 
components of individual DLA galaxies. This would also provide impact
parameters for absorption, which, from the measured incidence $dn/dz$, 
would allow the calculation of the space density (M\o ller \& Warren 
1998). The combination of information on both the stellar (emission) and 
interstellar (absorption) components of the DLA galaxies is crucial for
a complete understanding of these objects. At present, there is much 
debate about the nature of the DLA galaxies: on the basis of the velocity 
structure in metal absorption lines, Prochaska \& Wolfe (1997) argue that 
DLA absorbers are the fully-formed gaseous disks of present-day spiral 
galaxies, while Haehnelt et al. (1998) demonstrate that the same data
can be interpreted equally well as the signatures of merging protogalactic 
clumps. Helping to resolve such issues is the motivation for the imaging 
campaign described here.

In this paper we are concerned with DLA absorbers at high-redshift
$z>1.75$. (The best imaging study of low-redshift DLA absorbers is by
Le Brun et al 1997.) Attempts to image high-redshift DLA absorbers
from the ground have mostly been unsuccessful.  M\o ller and Warren
(1998) and Fynbo, M\o ller, and Warren (1999) provide a summary of the
observations and the conclusions that can be drawn from them. The
measured impact parameters of the handful of detections are in the
range $0.9\arcsec$ to $2.9\arcsec$, and magnitudes are $V\sim 25$. (At
$z=2.5$ an impact parameter of 1\arcsec corresponds to $3.9h^{-1}$ kpc
physical separation for $\Omega_m=1.0$, $\Omega_\Lambda=0.0$, and to
$5.7h^{-1}$ kpc for $\Omega_m=0.3$, $\Omega_\Lambda=0.7$, where
$h=H_{\circ}/100$.) One may reasonably suppose that the absorbers that
have escaped detection have similar impact parameters, but are
fainter, or have smaller impact parameters. In either case the clear
choice of telescope for a new imaging campaign is the Hubble Space
Telescope, for the unrivaled depth and spatial resolution
achievable\footnote{Bunker et al (1999) discuss the effectiveness of
blind spectroscopy for detecting emission from DLA absorbers, but
conclude that this is unlikely to be efficient even with an 8m
telescope.}.

We are engaged in a programme of imaging with the STIS and NICMOS
(NIC2) instruments aboard HST, to search for the galaxy counterparts
of 18 high-redshift $z>1.75$ DLA absorption lines seen in the spectra
of 16 target quasars.  This paper presents the results of the imaging
campaign with the NIC2 camera. The STIS images reach about 2.5 AB
mag. deeper than the NICMOS images, and have twice the resolution, and
therefore are expected to be the most useful in terms of proportion of
detections. The NICMOS observations complement the STIS observations,
providing colours, as well as luminosity profiles in the restframe
optical. Also the NICMOS frames will be more sensitive for any galaxy
counterparts that are particularly red. In Section 2 we tabulate the
target quasars, and the details of the observations, and describe the
steps followed in reducing the data and combining in mosaics. We
describe the methods used for subtracting the image of the quasar in
each field, and for constructing error frames that include the
systematic errors associated with the point spread function (psf)
subtraction. Section 3 includes a description of the detection
algorithm and catalogues the candidate galaxy counterparts found in
each field, together with details of their shapes, sizes, and
luminosity profiles, where measurable. The detection limits as a
function of angular separation from the quasar are also provided for
each field.

In a recent preprint Kulkarni et al. (2000) present similar NICMOS
observations of a single quasar field, where they detect a candidate
DLA absorber counterpart at small impact parameter,
$0.25\arcsec$. Their method of psf subtraction is different from
ours. Their preprint appeared after we had completed this work, and
the techniques were developed independently. We compare their method
of psf subtraction with ours, and discuss their detection in the light
of the detections reported here.

\section[]{Targets, observations, data reduction, mosaic combination, and psf
subtraction} 

\subsection{Targets and observations}

The target absorbers were selected with a view to covering a wide
range of column densities in order to be able to investigate any
correlation between column density and, for example, impact
parameter. We wished also to cover a broad redshift interval, to
quantify any evolution in the properties of the absorbing galaxies. A
lower redshift limit $z_{abs}=1.75$ was imposed as this corresponds to
the wavelength at which Ly$\alpha$ enters the range observable from
the ground. In fact many of the nearly 100 DLA absorbers so far
catalogued lie within quite a narrow range of redshifts $1.8<z<2.5$
(50$\%$ of the 80 DLA absorbers listed by Wolfe et al 1995).
A secondary consideration in selecting targets
was the desire to avoid very bright quasars, since the faint galaxy
counterparts would then be harder to detect in the glare of the quasar
light. 

We have imaged the fields around 16 quasars. In Table 1 we list in
successive columns: 1. quasar number $-$ this number is used
throughout this paper in referring to the fields by number, 2. quasar
name $-$ we have used the quasar name listed in the catalogue of
V\'{e}ron-Cetty and V\'{e}ron (1998), 3. statement of whether or not
the quasar has been detected at radio wavelengths, and 4. quasar
redshift $z_{em}$. In columns 5 and 6 are listed the column densities
${\mathrm N_{HI}}$ and redshifts $z_{abs}$ of all the catalogued
absorption lines of column density ${\mathrm N_{HI}}\,\ge 10^{20}$
cm$^{-2}$. Column 7 lists the references, in order, from which the
details listed in respectively columns 4, 5, and 6 were taken.  In the
final column of Table 1 each absorber is classified as DLA if the
column density is $>2\times 10^{20}$ cm$^{-2}$, and otherwise as LLS,
standing for Lyman-limit system. There are 18 absorbers classified
DLA, 5 classified LLS, and one candidate high-column density system
that remains to be confirmed. These are the absorbers for which we aim
to detect the counterpart galaxies.  In Fig. 1 we plot column density
against redshift for the 23 absorbers in Table 1 that have been
confirmed with high-resolution spectroscopy.

\begin{figure}
\vspace{9.0cm}
\includegraphics{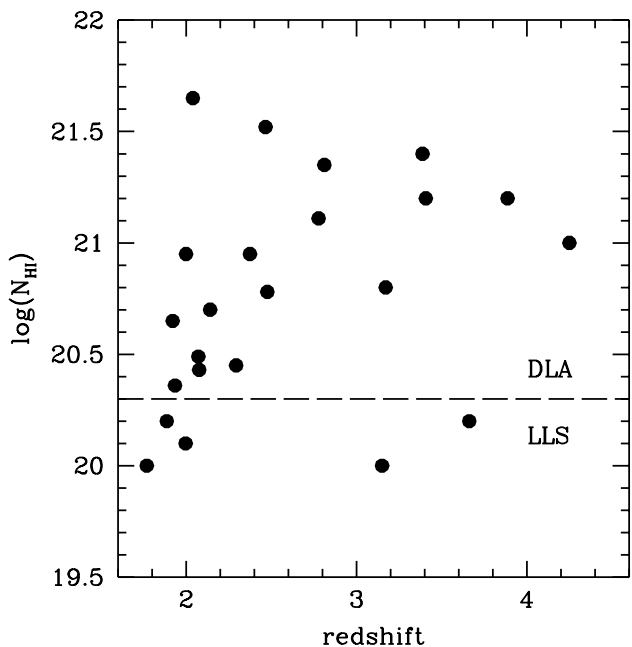}
\caption{Plot showing coverage of the column-density $-$ redshift
plane for the 23 confirmed absorbers listed in Table 1, 18 DLA 
absorption lines and 5 Lyman-limit systems.}
\end{figure}

\begin{table*}
  \caption{Redshifts and column densities of target absorption lines}
  \begin{tabular}{rlcllllc}
Quasar & Quasar & Radio & $z_{em}$ & $z_{abs}$ & ${\mathrm log({\mathrm
  N_{HI}}})$ & Refs & Classification \\ 
no.   & name & detected &          &          & ${\mathrm cm^{-2}}$&   &   \\
\hline \hline
 1  & CS\,73          & N & 2.256 & 1.8862 & 20.20 & 13, 11, 11 & LLS \\
    &                 &   &       & 2.0713 & 20.49 & $-$, 11, 11  & DLA \\
 2  & PC\,0056+0125   & Y & 3.154 & 2.7771 & 21.11 & 13, 12, 12 & DLA \\
 3  & PHL\,1222       & N & 1.922 & 1.9342 & 20.36 & 8, 8, 8    & DLA \\
 4  & PKS\,0201+113   & Y & 3.56  & 3.3875 & 21.4  & 13,16,16    & DLA \\
 5  & 0216+0803       & N & 2.992 & 1.7688 & 20.0: & 6, 6, 4    & LLS \\
    &                 &   &       & 2.2930 & 20.45 & $-$, 5, 5    & DLA \\
 6  & PKS\,0458$-$02  & Y & 2.286 & 2.0395 & 21.65 & 13, 11, 11 & DLA \\
 7  & PKS\,0528$-$250 & Y & 2.797 & 2.8110 & 21.35 & 1, 6, 9    & DLA \\
    &                 &   &       & 2.1408 & 20.75 & $-$, 10, 17  & DLA \\
 8  & H\,0841+1256    & N & 2.5:  & 1.86:  & ...   & 3, $-$, $-$    &
  unconfirmed$^a$ \\	      
    &                 &   &       & 2.3745 & 20.95 & $-$, 12, 12  & DLA \\
    &                 &   &       & 2.4764 & 20.78 & $-$, 12, 12  & DLA \\
 9  &PC\,0953+4749$^b$& N & 4.457 & 3.407  & 21.2  & 13, 2, 2   & DLA \\
    &                 &   &       & 3.887  & 21.2  & $-$, 2, 2    & DLA \\
    &                 &   &       & 4.250  & 21.0  & $-$, 2, 2    & DLA \\
10  & B2\,1215+33     & Y & 2.605 & 1.9990 & 20.95 & 13, 11, 11 & DLA \\
11  & Q\,1223+1753    & N & 2.936 & 2.4658 & 21.52 & 13, 11, 11 & DLA \\
12  & H\,1500$\#$13   & N & 3.249 & 3.1714 & 20.80 & 13,14,14  & DLA \\
13  & Q\,2116$-$358   & N & 2.340 & 1.9966 & 20.15 & 13,15,15  & LLS \\
14  & Q\,2206$-$1958  & N & 2.559 & 1.9205 & 20.65 & 13, 11, 11 & DLA \\
    &                 &   &       & 2.0762 & 20.43 & $-$, 11, 11  & DLA \\
15  & BR\,2212$-$1626 & N & 3.992 & 3.6617 & 20.20 & 6, 6, 6    & LLS \\
16  & 2233.9+1318     & N & 3.298 & 3.1501 & 20.00 & 13, 7, 7   & LLS \\ \hline
\end{tabular}
\begin{minipage}{170mm}
$^a$ A strong absorption line is seen near 3480\AA\, in a low-resolution
spectrum, which remains to be confirmed as Ly$\alpha$ with a
high-resolution spectrum (Hazard, private communication). \\
$^b$ The quoted absorption redshifts and column densities are preliminary.\\
Refs: [1] Bunker et al. 1999, [2] Bunker et al., in preparation, [3]
Hazard and Sargent, in preparation, [4] Lanzetta et al 1991, [5] Lu and
Wolfe 1994, [6] Lu et al 1996, [7] Lu et al 1993, [8] M\o ller et al
1998, [9] M\o ller and Warren 1993, [10] Morton et al 1980, [11]
Pettini et al 1994, [12] Pettini et al 1997, [13] V\'{e}ron-Cetty and
V\'eron 1998, [14] M\o ller, in preparation, [15] M\o ller et
al. 1994, [16] White et al. 1993, [17] Le Doux et al 1998.

\end{minipage}
\end{table*}

We used the NIC2 camera with the F160W (i.e. H band) filter to observe
the fields around these 16 quasars, at various dates commencing 23
April 1998 and ending 12 Sep 1998. The choice of the NIC2 camera was
arrived at through consideration of three factors: depth achieved,
sampling of the psf, and field of view. For NIC1 the smaller pixel
size carries with it a large read-out noise penalty, while the NIC3
camera pixel size is too large to allow accurate subtraction of the
psf. With NIC2 the psf is almost critically sampled, and the field of
view is large enough to allow the image to be stepped around the
array, for improved sky subtraction. 

Each field was observed for three orbits, and each orbit was split
into two exposures of length 1280 sec, resulting in six exposures per
quasar, totalling 7680 sec. The read-out mode employed was the
STEP-256 MULTIACCUM sequence. We used a rectangular dither pattern,
with a step size of $3\arcsec$, placing the quasar in the NIC2-FIX
aperture for the first exposure of each sequence. The HST Programme
Number of these observations is 7824. In Table 2 we list in successive
columns: 1. the quasar number from Table 1 (this is the Progamme Visit
Number in the HST data archive), 2. quasar name, 3. common
alternatives for the quasar name encountered in the literature,
4. J2000 coordinates measured from the Digital Sky Survey (DSS)
plates, 5. DSS plate used for the coordinates measurement, 6. The
F160W ${\mathrm H_{AB}}$ magnitude of the quasar measured from our data $-$
unless stated otherwise all magnitudes quoted in this paper are on the
AB system, 7. UT date at the start of the observations, 8. reference
number for locating the data in the HST archive. The target
PC\,0953+4749 is not visible on the DSS plate. To calculate the
position given in Table 2 we measured the coordinates of a nearby
galaxy on the DSS, and the offset to the quasar from the finding chart
in Schneider, Schmidt, and Gunn (1991).  As it happened HST was unable
to acquire the guide star for this target, so the frames are badly
trailed and could not be used. All the other observations were
successful.

\begin{table*}
  \caption{Details of target quasars and observations}
  \begin{tabular}{rllrrrrrrlclrrc}
\multicolumn{1}{c}{Quasar} & Quasar & Alternative & 
\multicolumn{3}{c}{Right Ascension} &
\multicolumn{3}{c}{Declination} & DSS & ${\mathrm H_{AB}}$ & 
\multicolumn{3}{c}{Date} & Archive \\
no.  & name & names & \multicolumn{6}{c}{(J2000)} &plate no. &(total) &
\multicolumn{3}{c}{observed} & ref. \\ \hline \hline
 1$\:\:$& CS\,73          & Q\,0049$-$2820 &  0 & 51 & 27.18 &
 $-$28 &  4 & 34.1 & 025A & 17.79 & Aug & 4  & 1998 & N4M701010 \\
 2$\:\:$& PC\,0056+0125   & Q\,0056+0125   &  0 & 59 & 17.57 &
   1   & 42 &  5.6 & 023A & 17.86 & Jul & 27 & 1998 & N4M702010 \\
        &                 & Q\,0056+014    &    &    &       &
       &    &      &      &       &     &    &      &           \\
 3$\:\:$& PHL\,1222       & Q\,0151+0448A  &  1 & 53 & 53.90 &
   5   &  2 & 57.1 & 04YF & 17.57 & Sep & 12 & 1998 & N4M703010 \\
        &                 & UM\,144        &    &    &       &
       &    &      &      &       &     &    &      &           \\
 4$\:\:$& PKS\,0201+113   & Q\,0201+1120   &  2 &  3 & 46.77 &
  11   & 34 & 46.2 & 03QO & 18.72 & Jul & 30 & 1998 & N4M704010 \\
 5$\:\:$& 0216+0803       & Q\,0216+0803   &  2 & 18 & 57.32 &
   8   & 17 & 27.8 & 002L & 17.60 & Sep & 12 & 1998 & N4M705010 \\
 6$\:\:$& PKS\,0458$-$02  & Q\,0458$-$0203 &  5 &  1 & 12.66 &
  $-$1 & 59 & 13.9 & 02O7 & 18.45 & Jul & 23 & 1998 & N4M706010 \\
 7$\:\:$& PKS\,0528$-$250 & Q\,0528$-$2505 &  5 & 30 &  7.94 &
 $-$25 &  3 & 30.1 & 04NM & 17.27 & Jul & 20 & 1998 & N4M707010 \\
 8$\:\:$& H\,0841+1256    &  0841+129      &  8 & 44 & 24.29 &
  12   & 45 & 47.0 & 028I & 17.75 & May & 13 & 1998 & N4M708010 \\
$9^a$   & PC\,0953+4749   & Q\,0953+4749   &  9 & 56 & 25.30 &
  47   & 34 & 42.5 & 00T8 & ...   & May & 15 & 1998 & N4M709010 \\
10$\:\:$& B2\,1215+33     & Q\,1215+3322   & 12 & 17 & 32.53 &
  33   &  5 & 38.1 & 00Y2 & 18.18 & Jul & 22 & 1998 & N4M710010 \\
        &                 & Q\,1215+333    &    &    &       &
       &    &      &      &       &     &    &      &           \\
11$\:\:$& Q\,1223+1753    &                & 12 & 26 &  7.19 &
  17   & 36 & 50.2 & 00G0 & 17.40 & Jul & 4  & 1998 & N4M711010 \\
12$\:\:$& H\,1500$\#$13   & Q\,1451+1223   & 14 & 54 & 18.55 &
  12   & 10 & 54.5 & 00J0 & 18.54 & Aug & 20 & 1998 & N4M712010 \\
13$\:\:$& Q\,2116$-$358   & Q\,2116$-$3550 & 21 & 19 & 27.58 &
 $-$35 & 37 & 41.2 & 02AD & 16.86 & Apr & 23 & 1998 & N4M713010 \\
14$\:\:$& Q\,2206$-$1958  & Q\,2206$-$199N & 22 &  8 & 52.07 &
 $-$19 & 43 & 59.5 & 020N & 16.81 & Aug & 9  & 1998 & N4M714010 \\
15$\:\:$& BR\,2212$-$1626 &                & 22 & 15 & 27.30 &
 $-$16 & 11 & 32.4 & 03YO & 17.70 & Aug & 5  & 1998 & N4M715010 \\
16$\:\:$& 2233.9+1318     & Q\,2233+1310   & 22 & 36 & 19.21 &
  13   & 26 & 21.0 & 02CV & 17.90 & Aug & 7  & 1998 & N4M716010 \\ 
\hline
\end{tabular}
\begin{minipage}{145mm}
\hspace{-1.5cm}$^a$ Guide star acquisition failed for this target, and
the images are too badly trailed to be useful.
\end{minipage}
\end{table*}

\subsection{Data reduction, and mosaic combination}

\subsubsection{Data reduction}

We used the STSDAS task {\bf calnica}, version 3.2, in the {\bf
nicmos} package to reduce the data, in conjunction with the new tasks
{\bf biaseq} and {\bf pedsky} in the {\bf nicproto} package, and using
the most recent calibration frames. NICMOS data suffer from pedestal
variations which produce offsets to the count level in each array
quadrant. Without correction the final frames will bear the imprint of
the flat-field frame, either negative or positive depending on the
pedestal level.  The tasks in the {\bf nicproto} package are designed
to remove this effect. The {\bf calnica} task includes routines for
subtracting the dark current, linearising the data, examining the
multiple reads to eliminate cosmic rays, and flat fielding the
data. Inspection of the data in the multiple reads for an individual
pixel is possible using the {\bf pstat} task, and this was used to
help in optimising the parameters of the different routines.  In some
cases vertical banding was visible in the final frames. These were
removed by subtracting from each column the median pixel value for
that column. The result of the first stage of the data reduction was
96 frames, comprising 6 exposures for each of the 16 quasars.

\begin{figure}
\vspace{7.7cm}
\includegraphics{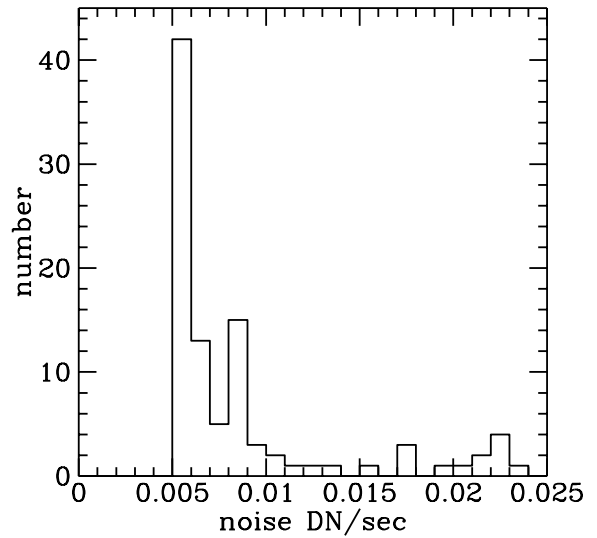}
\caption{Plot illustrating the variable quality of the data. The
histogram shows the distribution of measured values of the noise in
the sky in the 96 reduced frame, in units of DN sec$^{-1}$. Taking into
account read noise, dark noise, and Poisson noise from the counts in
the sky, the expected noise is in the range $0.005 - 0.006$ DN
sec$^{-1}$, corresponding to the peak in the histogram. As is evident,
less than half the data lie within this range.}
\end{figure}

The next step was to use the data frames themselves to remove
residuals, whether multiplicative (``flat field'') or additive (``sky
subtraction''). To a first approximation the sky level is the same in
all frames so that the residuals, from whatever cause, will be similar
in each frame and can be removed by subtracting the median of the 96
frames. The step size between pointings, $3\arcsec$, was chosen to be
sufficiently large that the quasar images from different frames would
not overlap, so that objects would be successfully removed in taking
the median.  Subtracting the median of the 96 frames resulted in a
very noticeable reduction in the noise in each frame.  We understand
that the explanation for this is that the frame used for dark
subtraction has substantial read noise, so that noise is being added
to each frame at the stage of dark subtraction. We remove this noise
by subtracting the median frame. There is a new task which produces
synthetic darks of low noise, which therefore avoids this
problem. However we were advised by the NICMOS group at STScI that the
procedure we followed produces equally satisfactory results. \footnote
{We investigated a second refinement to the removal of residuals, by
considering only the six frames for an individual quasar, and for each
frame subtracting the median of the other five frames. This led to an
increase in the noise in the background, and therefore was not
implemented.}

At this stage we created a list of bad pixels by identifying
discrepant high and low pixels in the median of the 96 frames, as well
as pixels where the recorded values showed a large dispersion in the
stack of 96 frames. The bad pixels were recorded as a ``mask'' frame
for use at the stage of combining the six data frames from each field.

The quality of the NICMOS data is highly variable. First there are a
few frames which contain an excessive number of cosmic ray hits. A
related but more serious problem is a consequence of the passage of
HST through the South Atlantic Anomaly (SAA) and the fact that these
HgCdTe arrays suffer from remanence i.e. very high counts in a pixel
are not completely flushed when the array is reset. Although the
NICMOS cameras are switched off during passage through the SAA, the
arrays experience a storm of cosmic-ray hits. In most cases the result
of remanence of the cosmic-ray hits is a background with a very high
noise level that affects subsequent images, and declines with
time. The remanence images of individual cosmic-ray hits are usually
difficult to discern, presumably indicating that the whole array has
been hit. To quantify the effect on the data we measured the standard
deviation in the counts in the background in each frame. Fig. 2 plots
the histogram of these noise measurements. The peak in the histogram
marks the noise in a clean frame, unaffected by the SAA. As shown in
Fig. 2 in $20\%$ of our frames the noise is more than double that in
an unadulterated frame and only about half the data are unaffected.

The NICMOS pipeline produces error frames which assume a Poisson noise
model. For the majority of frames this will be inappropriate because
of the additional source of noise discussed above. We therefore
created our own noise frames by adding in quadrature the Poisson
contribution from counts above sky to the measured noise in the
background.

\subsubsection{Mosaic combination}

The pipeline task {\bf calnicb} combines frames processed by {\bf
calnica} to produce a final mosaic image. We did not use {\bf calnicb}
because we wished to make a number of changes to the standard
pipeline. The relatively poor quality of a significant fraction of the
data demanded a more flexible approach that allowed each step in the
combination process to be scrutinised. We were particularly concerned
with identifying bad data, i.e. pixels with discrepant counts, either high
or low, from whatever cause: for example due to cosmic rays not
identified by {\bf calnica}, or from transient variation of the
quantum efficiency or dark current of any pixels. We used the routine
{\bf driz\_cr} in the {\bf ditherII} package (Fruchter and Hook 2000)
to identify bad data.  The first step was to align the six frames
using integer pixel offsets, and to form the median, ignoring pixels
in the bad-pixel mask. Each frame was then compared with the
combination frame to identify discrepant data values. The bad pixels
identified were combined with the original bad-pixel mask to form new
individual masks for each frame. A defect in the form of horizontal
bars is visible in some frames. These were identified and the pixels
added to the respective masks. Using the new masks a revised
combination frame was formed, and a revised identification of bad data
was made.

The step size between pointings, $3\arcsec$, is by design an integer
multiple of the pixel size, $0\farcs075$. The data frames were aligned
to the nearest pixel, and then averaged, weighting by the inverse of
the measured variance in the sky, and ignoring masked pixels. The
associated error frame was also created as appropriate for the
weighting and mask scheme used. We could have chosen sub-pixel dither
steps and to {\it drizzle} the data (Fruchter and Hook 2000). We chose
not to because the number of pointings is so small. In the event, in a
few frames the step pattern executed produced residuals from the
requested 40-pixel step of up to half a pixel.  The sub-pixel
residuals therefore broaden the psf very slightly. Fortunately the
residuals were almost identical for each target, so the psfs of each
quasar were affected in the same way.

The final frames are displayed in the left hand portion of Figs $5-8$ as
$S/N$ maps i.e. the image frame divided by the error frame. Displayed in
this way the reduced $S/N$ near the edges of the mosaic frames, where
the number of overlapping frames is smaller, is evinced by the smaller
number of sources visible on average in these regions. Over the
regions where there are only two overlapping frames the data are of
little value. This is because the scheme used to identify bad pixels
is ineffective if the number of frames is less than three.  In each of
the pictures the bright point source below and to the right of the
centre of the frame is the target quasar. In a number of cases objects
are visible at small angular separations from the quasar. These are
candidate counterparts of the DLA absorbers.

\subsection{Psf subtraction}

To identify candidates at small angular separation from the quasars,
and to measure their properties, we need to subtract the images of the
quasars. There are three possible methods for psf subtraction. We
could use the {\it Tiny Tim} software to create artificial psfs, or we
could use images of bright stars in the archive, or we could use the
15 quasar images themselves. The advantages of {\it Tiny Tim} psfs
over real psfs are that the psfs are noiseless, and can be created at
the required {\it x, y} location on the array (the psf varies with
{\it x, y} location), and at the precise sub-pixel position (which is
desirable as the psf is slightly undersampled). However because the
accuracy of the {\it Tiny Tim} psfs is limited by the detail in the
models, real psfs are preferred if suitably placed high $S/N$ images
are available. \footnote{After completing this work a new version of
{\it Tiny Tim} (5.0) was released, including modelling of the temporal
variation of the NICMOS psf, and other improvements. We have found
that the results obtained using the newest version of {\it Tiny Tim}
are as good as our results obtained using the quasar images
themselves, as quantified by measuring the dispersion in the
residuals. In some cases the {\it Tiny Tim} psfs are somewhat better,
and in some cases the quasar psfs are better. As suggested in the {\it
Tiny Tim} manual, this is probably because there are small,
essentially random, temporal variations in the psfs, so that, by
chance, one or the other method works better for a particular quasar.}

The main advantage of using images of bright stars is the high $S/N$,
with the proviso that any non-linear behaviour of the detector must be
accurately characterised. The undersampling of the psf nevertheless
sets a fundamental limit to the accuracy achievable if a single star
is used. Systematic residuals are unavoidable unless the star was
positioned at exactly the same sub-pixel position as the target. Other
disadvantages include possible poor colour match, and mismatch of the
psf in terms of {\it x, y} location. Primarily to deal with the
problem of undersampling we chose to use the 15 quasar images
themselves to subtract the psfs. Because we have 15 targets we can use
dithering to recover full sampling of the psf, without seriously
compromising the $S/N$ of the composite psf. Additional advantages of
using the quasars are that the colours are perfectly matched (assuming
the range in colour is small), and that the initial {\it x, y}
location and step pattern is the same for all the quasars. In addition
all the observations were completed over a few months, so the
condition of the telescope and camera was similar for each target. A
drawback to this approach is that it diminishes the possibility of
detecting the quasar host galaxies, since we will subtract the average
host-galaxy profile from each image.

The procedure followed for each quasar was to create a subtraction psf
by averaging the images of the other 14 quasars. Taking the relevant
14 images, the first step was to scale the images to the same count
level, and to register them to the nearest half pixel. Then, using the
{\bf dither} package, the data were interlaced into pixels of side
half the original size\footnote{In {\bf dither} parlance interlacing
corresponds to a delta-function {\it drizzle} footprint. Any
finite-sized footprint broadens the psf}. This sub-pixelised psf was
then scaled and registered to the quasar image by minimising $\chi^2$,
computing the flux at the location of a pixel by interpolation. In
computing the $\chi^2$, pixels within 0.3\arcsec of the quasar
centroid were excluded from the sum. Having subtracted the images of
all 15 quasars, mask frames were created by adding to the mask any
galaxies found close to the quasar, as well as any remaining bad
pixels visible. The whole procedure was then iterated, using the masks
to exclude these pixels in creating the psfs, and in making the
fits. In this second stage the quasar PHL\,1222 was excluding from the
averaging in creating subtraction psfs because of excess emission
clearly visible. This excess emission is centred on the quasar and is
likely to be from the host galaxy of the quasar.

Sub-pixelisation of the psf has two effects. First it improves the
sampling and so reduces systematic errors in subtraction near the
centre of the psf. However the resulting psf is noisier. Because of
the limited number of quasars used the psf subtraction adds noise to
the frames. This is more noticeable for the brightest quasars,
since the subtraction psf is formed by scaling up the images of
fainter quasars. The procedure could be improved, therefore, by using
a sub-pixelised psf for subtraction near the centre of the quasar
image, but using full-sized pixels away from the centre where the
issue of sampling is less important.

The approach Kulkarni et al. (2000) followed was to use single bright
stars for psf subtraction. They provide an extensive and detailed
analysis of the sources of uncertainty using this approach. In section
3.2.3 we compare the accuracy of our psf subtraction with that of
Kulkarni et al.

\subsection{Calculation of error frames}

\begin{figure*}
\vspace{9.5cm}
\includegraphics{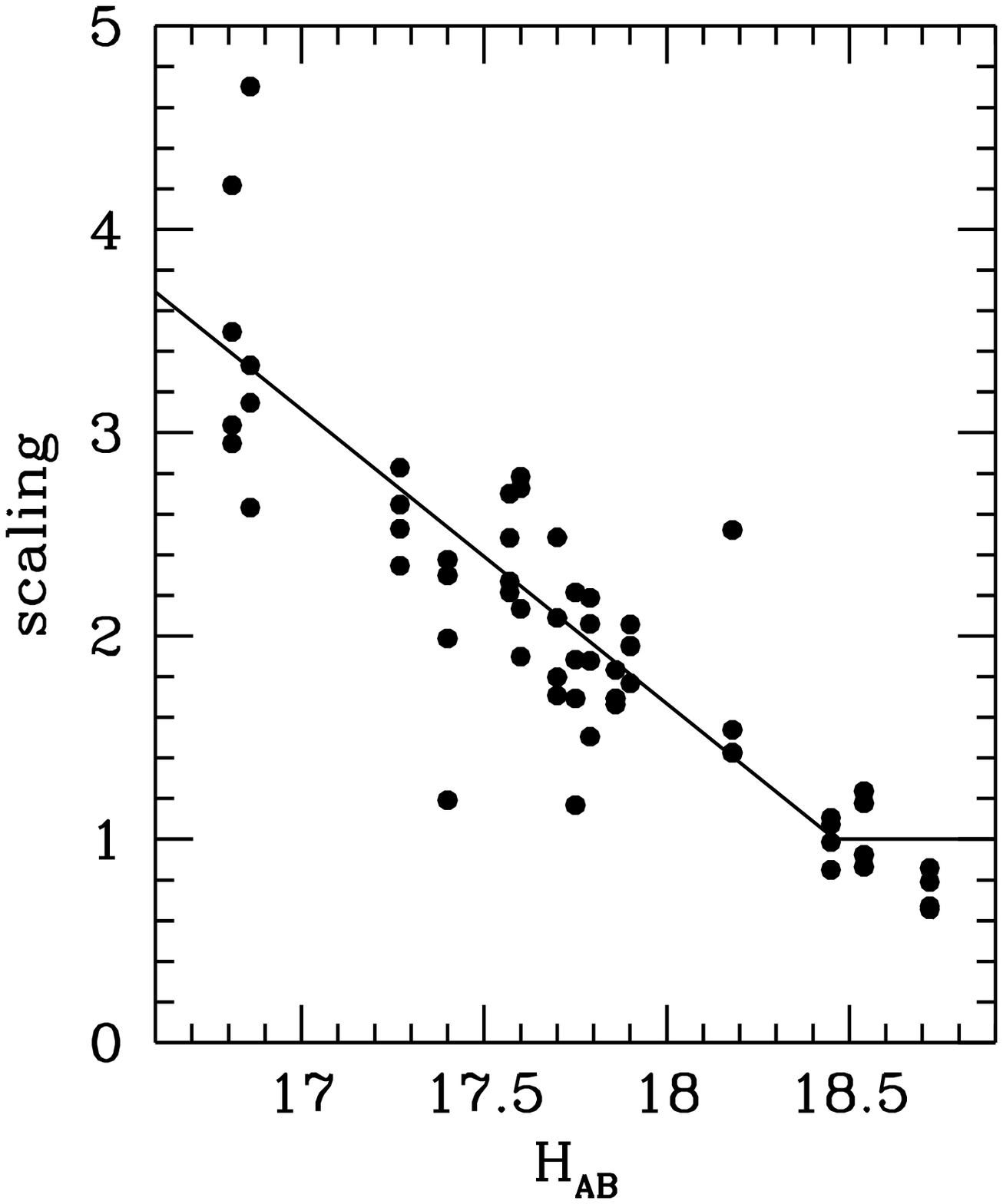}
\includegraphics{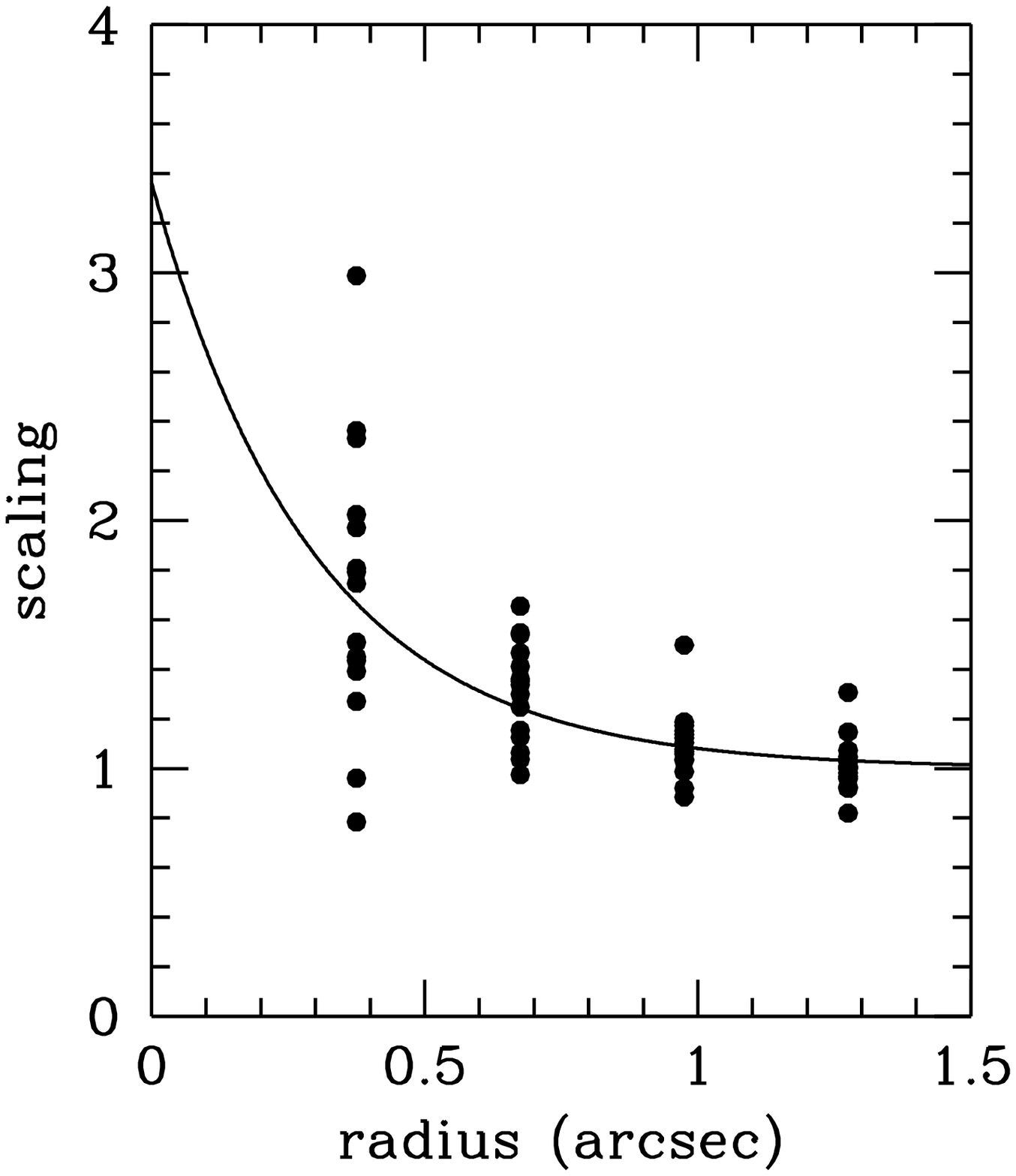}
\caption{Illustration of the rescaling of the Poisson errors to
account for systematic errors associated with the psf subtraction. At
each radius for each quasar the standard deviation $\sigma_{S/N}$ in
the $S/N$ frame is measured. A function of the form
$1.0+e^{-r/R}(a-b{\mathrm H_{AB}})$ is fit to the data. To plot all
the data at once we apply the results of the fit to rescale the
measured values to a specific radius $r=5$ pixels ($0.375\arcsec$)
(left-hand plot), and a specific magnitude ${\mathrm H_{AB}=18.0}$
(right-hand plot). The left-hand plot then shows how the scaling
increases for brighter quasars, and the right-hand plot shows how the
scaling increases at smaller radii. The solid lines are the fitted
function.}
\end{figure*}

The final stage before searching for candidates was to compute revised
$1\sigma$ error frames for each quasar. The purpose of this is to be
able to measure the significance of candidate detections. The revised
error frames must include the contribution of random and systematic
errors associated with the psf subtraction. We first computed $S/N$
frames by dividing the subtracted frames by the current error
frames. Ideally, if the subtraction psf were both noiseless, and
perfectly accurate, then the distribution of counts in the $S/N$ frames,
outside remaining objects, would be Gaussian with standard deviation
$\sigma_{S/N}=1.0$. Because the subtraction psf is not noiseless, to
account for the added noise we scaled the error frame by the measured
$\sigma_{S/N}$ in the sky, and then created new $S/N$ frames. Even then
$\sigma_{S/N}$, although now unity in the sky, increased towards the
centre of the (subtracted) quasar. The increase was greater for the
brighter quasars. The reason for this can be understood as follows. For
bright objects, such as the quasars in this programme, the errors are
dominated by object photon noise i.e. $\sqrt N$, where $N$ is the number
of detected photons. Suppose that the psf has a fractional accuracy
$f$ i.e. the typical residual for any pixel after psf subtraction is
$fN$. Then in the $S/N$ frame the residuals will be visible as
deviations of $S/N=f\sqrt N$, i.e. larger where the counts
are larger, and larger for brighter quasars.

To quantify these systematic errors, for each quasar we measured
$\sigma_{S/N}$ in annuli centred on the quasar, clipping out real
objects. To these results we fit a function of the form
$\sigma_{S/N}={\mathrm max}(1.0+e^{-r/R}(a-b {\mathrm H_{AB}}), 1.0)$.
The error frames were then rescaled by the fitted function. In this
equation $r$ is the distance from the quasar centroid in pixels, $R$
is a scale length, $a$ and $b$ are constants, and ${\mathrm H_{AB}}$
is the apparent magnitude of the quasar. The fit of the function is
illustrated in two plots in Fig. 3.  These plots indicate that for
quasars fainter than ${\mathrm H_{AB}}=18.5$ it is possible to achieve
psf subtraction with negligible systematic error, but that the
systematic errors increase rapidly at brighter magnitudes.

It is noticeable from the right hand plot in Fig. 3 that the scatter
increases at smaller radii, so that the scaling becomes less reliable
near the centre. From an inspection of the residuals for the different
targets we judged that inside a radius of 4.5 pixels (0.34\arcsec) the
psf subtraction is unreliable, and we set the data to zero inside this
radius.

\section{Detection of candidates, and catalogues}

\begin{figure*}
\vspace{12cm}
\includegraphics{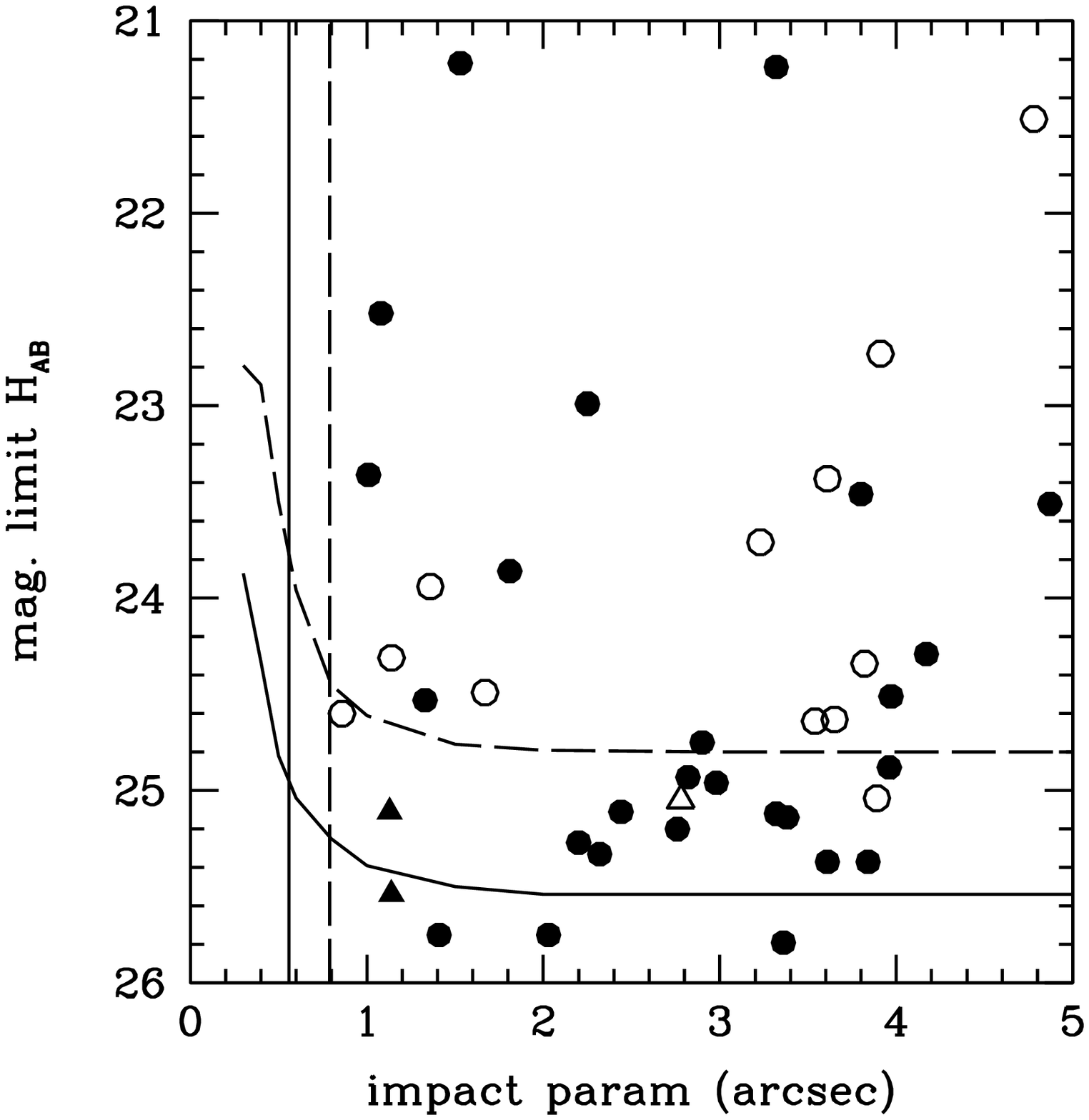}
\caption{Plot of relation between ${\mathrm H_{AB}}$
aperture magnitude and impact parameter for the 41 candidate DLA galaxies,
taken from Table 4. Solid symbols identify compact objects (C in Table
4), and open symbols diffuse objects (D in Table 4): circles mark
candidates, triangles mark sources confirmed as the DLA
counterpart. The curves mark the detection limits averaged for the 15
fields, {\it solid} for compact objects and {\it dashed} for diffuse
sources.  The vertical solid and dashed lines mark the inner radius
limit used for the search for objects, for the small aperture
(i.e. compact objects) and large aperture (i.e.diffuse objects)
respectively. Candidates with centroids inside these limits are not
considered reliable and have been excluded from the catalogue in Table
4.}
\end{figure*}

\subsection{Detection}

In this subsection we describe the algorithm used to produce a list of
candidates. We restricted the search to a box of side $7.5\arcsec$
centred on the quasar.  A commonly used procedure for detecting
sources is to convolve the frame with a filter, typically a Gaussian,
and to identify peaks in the convolved frame. There were three reasons
why we did not follow this approach. The first is that the noise in
the subtracted frames is larger near the quasar centre, so we need a
detection threshold that varies as the noise varies.  The second is
that objects with surface-brightness profiles very different from the
filter profile will not be detected. So, for example, low-surface
brightness galaxies could be missed. The third is that we wish to
quote detection limits in a way such that it is possible easily to
compute whether a hypothetical galaxy (of arbitrary shape, profile,
and magnitude) would have been detected.

Our method is to convolve with detection filters in a manner that
produces complete samples of objects that have $S/N$ above a specified
threshold, as measured within circular apertures of specified sizes.
The data and variance frames are convolved with circular top-hat
filters, of size equal to the requested aperture. The frame formed by
taking the ratio of the convolved data frame and the square root of
the convolved variance frame contains, at each pixel, the summed
$S/N$ within the aperture of any source centred at that pixel. This
ratio frame is then searched for peaks above a chosen $S/N$ threshold. 

We first ran a large-box median filter over the frames, and subtracted
this off, to remove any remaining gradual variations in the
background. We used two detection apertures, of diameter 6 pixels (0.45\arcsec)
and 12 pixels (0.90\arcsec). The smaller aperture was designed to
detect compact sources, and the larger aperture was designed to find
objects of lower surface-brightness that had been missed by the first
filter.  The $S/N$ detection threshold was selected from a consideration
of the distribution of the depths of negative peaks in the aperture
$S/N$ frames. With the exception of one field there were no negative
detections in any frame below $S/N=-6$ for either detection
filter. For the quasar Q\,$2116-358$ the psf subtraction was less
successful than for the other targets, and there are negative
residuals at the position of two of the diffraction spikes with
$S/N\sim -8$. With this in mind we set our detection threshold at
$S/N=6$ with the expectation that nearly all the objects detected will
be real, with the caveat that sources located on diffraction spikes
will be less reliable. In fact with the exception of very red objects
we expect that all real objects in our candidate list will be detected
at higher $S/N$ in our STIS data. Therefore we will reassess the
reliability of the candidate list on completion of the list of
candidates detected in the STIS frames.

{\it Inner radius limit.} Because the psf subtraction inside
$r=0.34\arcsec$ was judged unreliable (section 2.4) the inner radius to
which objects can be detected is equal to the sum of 0.34\arcsec and
the aperture radius. Therefore for the small aperture the inner radius
limit is 0.56\arcsec, and for the large aperture the inner radius
limit is 0.79\arcsec.

{\it Detection limits.} The candidate list, then, was constructed by
identifying any peaks higher than 6.0 in the aperture $S/N$ frames. Real
objects in the frames will be detected at higher $S/N$ with the filter
that most closely matches the galaxy surface brightness
profile. Therefore if a peak was found at the same location for both
filters we eliminated the detection of lower $S/N$, and classified the
object as compact (C) or diffuse (D) for detection with the small or
large aperture filter respectively. At the location of each peak we
integrated the counts within the aperture, and transformed this to an
aperture ${\mathrm H_{AB}}$ magnitude using the photometry zero point
provided in the file header (the data were calibrated by converting the
count rates (DN/sec) to Jy by mulitplying by the factor
$2.070\times10^{-6}$). Then for each frame and for both filters
we measured the average detection limit in annuli at several different
radii. These values are listed In Table 3. In Fig. 4 we plot, for
both filters, the radial profile of the detection limits averaged for
the 15 fields. The average detection limit for compact (diffuse)
sources is ${\mathrm H_{AB}}=25.0\, (24.4)$ at an angular separation
of $0.56\arcsec\, (0.79\arcsec)$ from the quasar, improving to
${\mathrm H_{AB}}=25.5\, (24.8)$ at large angular separations.

{\it Aperture magnitudes.} All the magnitudes of the candidates
(section 3.2) are quoted as aperture magnitudes i.e. we have simply
summed the counts within the aperture and no aperture correction has
been applied (aperture corrections are, nevertheless, discussed in
section 3.2.1). We have not attempted to extrapolate to total
magnitudes, since this would be unreliable for the faint sources. The
use of aperture magnitudes makes it very simple to compute whether or
not a hypothetical galaxy of specified surface brightness profile and
impact parameter would have been detected in a particular field. One
first computes the aperture magnitude of the galaxy, for the small and
large apertures. For the hypothesised impact parameter one compares
the two aperture magnitudes against the detection limits provided in
Table 3 for the field in question. If the galaxy were brighter than
either detection limit it would have been included in our catalogue.

\begin{table*}
  \caption{Aperture magnitude ${\mathrm H_{AB}}$ average detection limits as a
  function of impact parameter for each field, for compact and diffuse
  sources.} 
  \begin{tabular}{rlcccccccccc}
No. & Quasar & compact or & \multicolumn{9}{c}{impact parameter (arcsec)} \\
 & & diffuse & 0.3 & 0.4 & 0.5 & 0.6 & 0.8 & 1.0 & 1.5 & 2.0 & 3.0 \\
\hline \hline
 1& CS\,73          &C & 23.77 & 24.34 & 24.94 & 25.21 & 25.44 & 25.58 & 25.71 & 25.72 & 25.74 \\
  &               &D & 22.65 & 22.75 & 23.47 & 24.02 & 24.63 & 24.80 & 24.95 & 24.98 & 25.00 \\
 2& PC\,0056+0125   &C & 23.89 & 24.44 & 24.95 & 25.16 & 25.37 & 25.48 & 25.59 & 25.62 & 25.63 \\
  &               &D & 22.80 & 22.92 & 23.60 & 24.10 & 24.57 & 24.72 & 24.84 & 24.88 & 24.89 \\
 3& PHL\,1222       &C & 23.51 & 24.08 & 24.74 & 25.08 & 25.39 & 25.59 & 25.75 & 25.78 & 25.79 \\
  &               &D & 22.35 & 22.45 & 23.20 & 23.84 & 24.55 & 24.79 & 25.00 & 25.04 & 25.05 \\
 4& PKS\,0201+113   &C & 25.33 & 25.56 & 25.71 & 25.76 & 25.78 & 25.80 & 25.81 & 25.81 & 25.81 \\
  &               &D & 24.46 & 24.54 & 24.82 & 24.95 & 25.03 & 25.05 & 25.07 & 25.07 & 25.07 \\
 5& 0216+0803     &C & 23.42 & 24.02 & 24.71 & 25.03 & 25.31 & 25.49 & 25.64 & 25.69 & 25.70 \\
  &               &D & 22.23 & 22.34 & 23.13 & 23.75 & 24.48 & 24.71 & 24.89 & 24.94 & 24.96 \\
 6& PKS\,0458$-$02  &C & 25.24 & 25.39 & 25.51 & 25.54 & 25.55 & 25.56 & 25.57 & 25.57 & 25.57 \\
  &               &D & 24.36 & 24.45 & 24.66 & 24.75 & 24.80 & 24.80 & 24.82 & 24.83 & 24.83 \\
 7& PKS\,0528$-$250 &C & 23.13 & 23.70 & 24.33 & 24.65 & 24.95 & 25.16 & 25.33 & 25.39 & 25.40 \\
  &               &D & 21.97 & 22.07 & 22.81 & 23.41 & 24.12 & 24.37 & 24.59 & 24.64 & 24.66 \\
 8& H\,0841+1256    &C & 23.81 & 24.20 & 24.61 & 24.78 & 24.98 & 25.10 & 25.21 & 25.24 & 25.25 \\
  &               &D & 22.75 & 22.86 & 23.45 & 23.84 & 24.19 & 24.34 & 24.46 & 24.50 & 24.51 \\
10& B2\,1215+33     &C & 24.55 & 24.97 & 25.31 & 25.45 & 25.57 & 25.64 & 25.68 & 25.70 & 25.70 \\
  &               &D & 23.48 & 23.58 & 24.16 & 24.52 & 24.80 & 24.88 & 24.93 & 24.95 & 24.96 \\
11& Q\,1223+1753    &C & 23.25 & 23.77 & 24.29 & 24.55 & 24.82 & 25.00 & 25.16 & 25.20 & 25.21 \\
  &               &D & 22.15 & 22.26 & 22.94 & 23.47 & 24.00 & 24.21 & 24.41 & 24.46 & 24.47 \\
12& H\,1500$\#$13   &C & 25.27 & 25.49 & 25.65 & 25.69 & 25.70 & 25.71 & 25.72 & 25.72 & 25.72 \\
  &               &D & 24.41 & 24.48 & 24.75 & 24.88 & 24.95 & 24.97 & 24.98 & 24.98 & 24.98 \\
13& Q\,2116$-$358   &C & 22.72 & 23.27 & 23.86 & 24.22 & 24.57 & 24.82 & 25.05 & 25.11 & 25.12 \\
  &               &D & 21.46 & 21.56 & 22.40 & 22.97 & 23.72 & 24.01 & 24.31 & 24.36 & 24.38 \\
14& Q\,2206$-$1958  &C & 22.49 & 23.06 & 23.92 & 24.34 & 24.73 & 25.04 & 25.30 & 25.38 & 25.40 \\
  &               &D & 21.23 & 21.34 & 22.10 & 22.89 & 23.87 & 24.21 & 24.55 & 24.63 & 24.65 \\
15& BR\,2212$-$1626 &C & 23.65 & 24.11 & 24.58 & 24.76 & 24.96 & 25.09 & 25.22 & 25.25 & 25.25 \\
  &               &D & 22.61 & 22.71 & 23.36 & 23.77 & 24.17 & 24.33 & 24.46 & 24.50 & 24.51 \\
16& 2233.9+1318   &C & 24.02 & 24.56 & 25.14 & 25.39 & 25.60 & 25.73 & 25.83 & 25.85 & 25.86 \\
  &               &D & 22.88 & 22.99 & 23.68 & 24.26 & 24.81 & 24.95 & 25.08 & 25.11 &
25.12 \\ \hline 
\end{tabular}
\end{table*}

\begin{table*}
  \caption{Coordinates and aperture magnitudes of candidate DLA galaxies}
  \begin{tabular}{rllrrlrrlcrrrl}
\multicolumn{1}{c}{Quasar} & Quasar & Candidate & 
\multicolumn{3}{c}{Right Ascension} & \multicolumn{3}{c}{Declination} &
 ${\mathrm H_{AB}}$ & $S/N$ & Offset &\multicolumn{1}{c}{PA}& Remarks \\
no. & name & no. & \multicolumn{6}{c}{(J2000)} &(aper)& &
  \multicolumn{1}{c}{\arcsec} & $^{\circ}$ (E of N) &  \\ \hline \hline
 1&CS\,73         &       &     &       &   & & & & & & & & no candidates \\ \hline
 2&PC\,0056+0125  &N-2-1C & 0&59&17.7082&  1&42&06.684&25.33&  8.3&2.32&  62.1& \\
  &             &N-2-2C & 0&59&17.7468&  1&42&08.403&25.37&  7.8&3.84&  43.2& \\
  &             &N-2-3C & 0&59&17.4756&  1&42&09.344&24.51& 16.8&3.97& -20.4& \\
  &             &N-2-4D & 0&59&17.2594&  1&42&04.345&21.51&123.3&4.78&-105.3& \\ \hline
 3&PHL\,1222      &N-3-1C & 1&53&53.7802&  5&02&56.764&23.86& 33.5&1.81&-100.9& \\
  &             &N-3-2C & 1&53&54.0429&  5&02&59.652&21.24&248.6&3.32&  39.8& \\
  &             &N-3-3D & 1&53&53.6837&  5&02&55.440&23.38& 27.7&3.61&-117.4& \\ \hline
 4&PKS\,0201+113  &N-4-1C & 2&03&46.8345& 11&34&46.579&23.36& 53.6&1.01&  67.9& \\
  &             &N-4-2C & 2&03&46.9602& 11&34&45.318&24.75& 16.2&2.90& 107.4& \\
  &             &N-4-3C & 2&03&46.7978& 11&34&42.849&25.79&  6.6&3.36& 173.3& \\
  &             &N-4-4C & 2&03&46.8794& 11&34&42.548&24.88& 14.2&3.96& 156.5& \\ \hline
 5&0216+0803    &N-5-1C & 2&18&57.2276&  8&17&27.443&25.75&
  6.3&1.41&-104.8&Note 1 \\
  &             &N-5-2D & 2&18&57.5773&  8&17&27.264&24.34& 10.5&3.82&  97.8& \\ \hline
 6&PKS\,0458$-$02 &N-6-1D & 5&01&12.6411& -1&59&14.711&24.60&  7.6&0.86&-160.8& \\
  &             &N-6-2D & 5&01&12.6940& -1&59&15.502&24.49&  8.1&1.67& 162.5& \\
  &             &N-6-3C & 5&01&12.5480& -1&59&16.177&24.93& 11.6&2.82&-143.8& \\
  &             &N-6-4C & 5&01&12.8168& -1&59&12.045&24.96& 11.5&2.98&  51.5& \\
  &             &N-6-5C & 5&01&12.8944& -1&59&11.610&24.29& 19.9&4.17&  56.7& \\ \hline
 7&PKS\,0528$-$250&N-7-1C & 5&30&07.9705&-25&03&31.170&25.54&  5.0&1.14&
  159.0&Note 2 \\
  &             &N-7-2D & 5&30&08.0753&-25&03&33.271&24.63&  6.3&3.65& 150.1& \\ \hline
 8&H\,0841+1256   &N-8-1C & 8&44&24.2166& 12&45&47.182&22.52& 64.6&1.08& -80.6& \\
  &             &N-8-2C & 8&44&24.4526& 12&45&48.413&25.20&  6.7&2.76&  59.3& \\
  &             &N-8-3D & 8&44&24.3525& 12&45&43.156&22.73& 30.7&3.91& 166.8& \\ \hline
10&B2\,1215+33    &N-10-1C&12&17&32.5264& 33&05&36.052&25.75&  6.3&2.03&-178.5& \\
  &             &N-10-2D&12&17&32.4597& 33&05&34.975&23.71& 18.6&3.23&-164.0& \\ \hline
11&Q\,1223+1753   &       &     &       &   & & & & & & & & no candidates \\ \hline
12&H\,1500$\#$13  &N-12-1D&14&54&18.4720& 12&10&50.759&25.04&  6.2&3.89&-162.8& \\ \hline
13&Q\,2116$-$358  &N-13-1D&21&19&27.4733&-35&37&41.641&23.94&
  8.1&1.36&-109.0&Note 1 \\
  &             &N-13-2C&21&19&27.5516&-35&37&45.003&23.46& 27.7&3.80&-174.6& \\
  &             &N-13-3C&21&19&27.9659&-35&37&39.795&23.51& 27.0&4.87&  73.1& \\ \hline 
14&Q\,2206$-$1958 &N-14-1C&22&08&52.1019&-19&43&58.456&25.11& 6.5&1.13&
  23.6&Notes 1, 3 \\
  &             &N-14-2C&22&08&52.0480&-19&43&58.192&24.53& 12.1&1.33& -13.2& \\
  &             &N-14-3C&22&08&52.1966&-19&43&58.213&25.27&  7.2&2.20&  54.3& \\
  &             &N-14-4C&22&08&51.8986&-19&43&59.055&25.11&  7.8&2.44& -79.9& \\
  &             &N-14-5D&22&08&52.2106&-19&44&02.484&24.64&  6.3&3.54& 146.3& \\ \hline
15&BR\,2212$-$1626&N-15-1D&22&15&27.3782&-16&11&32.611&24.31&  6.7&1.14& 100.4& \\
  &             &N-15-2C&22&15&27.1966&-16&11&32.784&21.22&207.5&1.53&-104.7& \\
  &             &N-15-3C&22&15&27.3681&-16&11&34.450&22.99& 47.5&2.25& 154.6& \\
  &             &N-15-4C&22&15&27.5349&-16&11&32.041&25.14&  7.0&3.38&  83.7& \\
  &             &N-15-5C&22&15&27.1766&-16&11&29.211&25.37&  6.2&3.61& -29.0& \\ \hline
16&2233.9+1318  &N-16-1D&22&36&19.2800& 13&26&18.405&25.05&  6.6&2.78&
  158.6&Note 4 \\
  &             &N-16-2C&22&36&19.4239& 13&26&22.236&25.12& 12.2&3.32&
  68.3& \\ \hline
\end{tabular}
\begin{minipage}{170mm}
1. The three sources N-5-1C, N-13-1D, and N-14-1C lie on
diffraction spikes and should therefore be considered less certain/reliable
than the other sources. \\
2. This source is the DLA at $z=2.8110$ confirmed by M\o ller and
Warren (1993), and so has been included even though it has $S/N<6$. \\
3. This source has been confirmed by us as the DLA at $z=1.9205$ 
(M\o ller et al, in preparation). \\
4. This source is the LLS at $z=3.1501$ confirmed by Djorgovski et
al. (1996). 
\end{minipage}
\end{table*}

\subsection{Catalogue}

The catalogue of detections is provided in Table 4. For each quasar
field the candidates are listed in order of angular separation from
the quasar. Listed in the first three columns of Table 4 are the
quasar number, quasar name, and the candidate number.  The candidate
numbering scheme is explained as follows. Taking as example candidate
N-15-3C, the `N' stands for NICMOS (we will provide a similar list `S'
for the STIS images), 15 is the quasar number, the 3 indicates that it
is the third nearest candidate to the quasar in that field, and the C
states that it is compact. We find a total of 41 candidates. The
magnitudes and impact parameters are compared against the average
detection limits in Fig. 4. The positions of the candidates are
illustrated in the finding charts in the right-hand panels in Figs
$5-8$.  The remaining columns in Table 4 list successively the
coordinates, aperture magnitude, detection $S/N$, angular separation
from the quasar, and the position angle from the quasar. The NICMOS
data header files contain an astrometric solution that provides
coordinates that are accurate in a relative sense, for the small
angular separations with which we are concerned. The coordinates were
shifted to absolute values by adopting for the coordinates of the
quasar the value measured from the DSS plate, listed in Table 2.

\begin{table*}
  \caption{Sersic profile fits to brighter candidates}
  \begin{tabular}{rllccrcccl}
\multicolumn{1}{c}{No.} & Quasar & Candidate & $r_e$ & $n$ &
  \multicolumn{1}{c}{orient'n} & ellip & ${\mathrm H_{AB}}$ &
 ${\mathrm \Delta H_{AB}}$ & comments \\
 & & & $\arcsec$ & & $^{\circ}$ (E of N) & & (total) & (ap. corr.)\\ \hline \hline
 2&PC\,0056+0125  &N-2-4D &$0.212^{+0.008}_{-0.008}$&
$2.57^{+0.15}_{-0.14}$&$64.4^{+0.5}_{-0.5}$&$0.70^{+0.01}_{-0.01}$&21.06 &0.45 & \\
 3&PHL\,1222      &N-3-1C &$0.119^{+0.006}_{-0.006}$&
$0.65^{+0.17}_{-0.16}$&$-27.1^{+...}_{-...}$&$0.05^{+0.09}_{-...}$&23.35 &0.51 & Note 1 \\
  &               &N-3-2C & & & & & & & point source (Q\,0151+0448B) \\
  &               &N-3-3D &$0.437^{+0.053}_{-0.035}$&
$0.84^{+0.21}_{-0.16}$&$-27.2^{+11.5}_{-10.6}$&$0.21^{+0.07}_{-0.07}$&
22.54  &0.84 &   faint companions masked \\ 
 4&PKS\,0201+113  &N-4-1C &$0.203^{+0.011}_{-0.009}$&
$1.21^{+0.15}_{-0.13}$&$-1.9^{+3.0}_{-3.0}$&$0.38^{+0.03}_{-0.03}$&22.40 &0.96 &  \\
 6&PKS\,0458$-$02 &N-6-5C &$0.103^{+0.008}_{-0.008}$&
$0.41^{+0.25}_{-0.19}$&$53.0^{+4.5}_{-4.5}$&$0.60^{+0.06}_{-0.07}$&23.79 &0.50 & \\
 8&H\,0841+1256   &N-8-1C &$0.068^{+0.005}_{-0.005}$&
$2.69^{+0.50}_{-0.40}$&$-53.1^{+1.7}_{-1.7}$&$0.70^{+0.03}_{-0.03}$&21.95 &0.57 & \\
  &               &N-8-3D &$0.371^{+0.018}_{-0.015}$&
$0.61^{+0.12}_{-0.10}$&$-78.8^{+5.1}_{-4.9}$&$0.34^{+0.04}_{-0.05}$&22.11 &0.62 & \\
10&B2\,1215+33    &N-10-2D&$0.256^{+0.035}_{-0.024}$&
$0.93^{+0.36}_{-0.25}$&$72.4^{+2.5}_{-2.6}$&$0.73^{+0.04}_{-0.04}$&23.21 &0.50 & \\
13&Q\,2116$-$358  &N-13-2C&$0.130^{+0.026}_{-0.016}$&
$2.23^{+0.82}_{-0.54}$&$-86.0^{+13.4}_{-12.8}$&$0.28^{+0.09}_{-0.10}$&22.74 &0.72 & \\
  &               &N-13-3C&$0.170^{+0.010}_{-0.009}$&
$0.65^{+0.20}_{-0.16}$&$-29.7^{+9.0}_{-8.9}$&$0.28^{+0.07}_{-0.08}$&22.74 &0.77 & \\
15&BR\,2212$-$1626&N-15-2C& & & & & & & point source \\
  &               &N-15-3C&$0.115^{+0.006}_{-0.006}$&
$1.28^{+0.22}_{-0.19}$&$89.3^{+2.9}_{-2.9}$&$0.50^{+0.04}_{-0.04}$&22.37 &0.62 & 
  \\ \hline
\end{tabular}
\begin{minipage}{170mm}
1. This source is almost round. In consequence the confidence limits
   on the orientation are indeterminate, as is the lower confidence limit for
   the ellipticity.

\end{minipage}
\end{table*}

\subsubsection{Surface-brightness profiles}

For the brighter candidates we have measured the deconvolved surface
brightness profile using the techniques described by Warren et al
(1996). We use the Sersic model where the surface brightness as a
function of radius $r$ is $\Sigma=\Sigma_e
\exp\{-B(n)\lbrack(r/r_e)^{1/n}-1\rbrack\}$. The parameter $n$
characterises the shape of the profile: $n=1$ is the exponential
profile and $n=4$ the de Vaucouleurs profile. This parameterisation is
particularly useful, therefore, as the value of $n$ can be used to
classify faint galaxies into ``early'' and ``late'' types.  The parameter
$r_e$ is the half-light radius, $\Sigma_e$ is the surface brightness
at the half-light radius, and $B(n)$ is a constant for particular
$n$. Ciotti and Bertin (1999) provide the series asymptotic solution for
$B(n)$, and we used the approximation provided by the first four terms
$B(n)=2n-1/3+4/(405n)+46/(25515n^2)$ which is accurate to better than
one part in $10^6$ over the range of $n$ of interest.

We created a psf using the {\it Tiny Tim} software, with pixels
one-third the size of the NIC2 pixels, to ensure adequate sampling.
The fitting proceeds by creating a two-dimensional galaxy
surface-brightness profile with the same small pixel size, convolving
with the psf, rebinning to the full pixel size and computing the
$\chi^2$ of the fit. The best fit is found by $\chi^2$ minimisation on
the seven parameters $x$, $y$, $\Sigma_e$, $r_e$, $n$, orientation,
and ellipticity. In Table 5 we provide the details of the fits to all
the candidates in Table 4 with $S/N>18$. Also listed there are the
total magnitudes, computed by extrapolating the models to infinite
radius, and the aperture corrections i.e. the difference between the
total and aperture magnitudes.  It is of interest to compare the
aperture corrections for these galaxies with the values for a point
source. For a point source the aperture correction for the small
aperture (used for compact objects) is 0.444 mag.  Excluding the two
point sources listed in Table 5, there are seven compact objects, for
which the mean aperture correction is 0.66 mag., with a scatter of
0.16 mag.  For a point source the aperture correction for the large
aperture (used for diffuse objects) is 0.134 mag. There are four
diffuse objects listed in Table 5, for which the mean aperture
correction is $0.60\pm0.17$ mag.

\begin{figure*}
\vspace{20.8cm}
\includegraphics{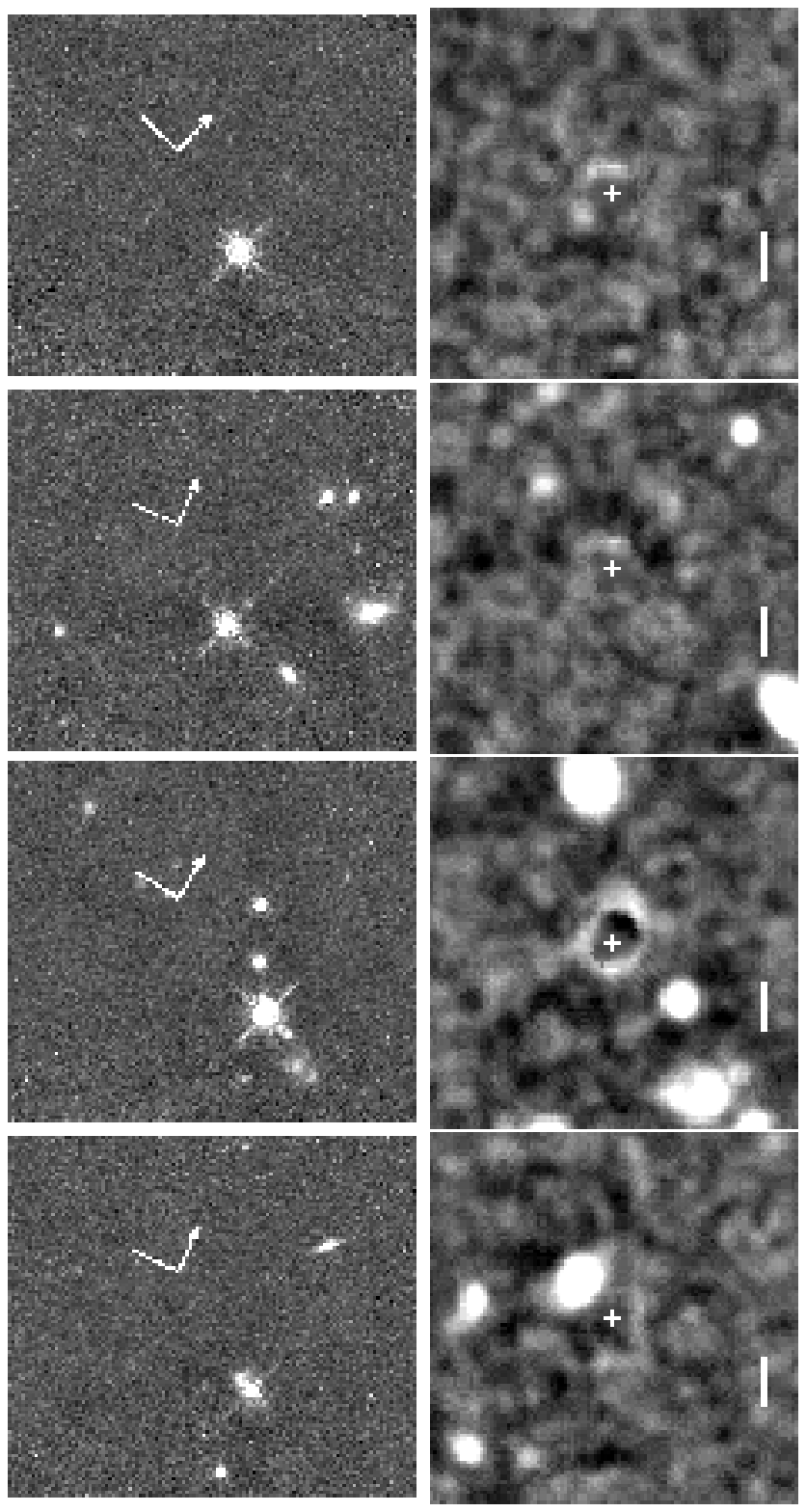}
\includegraphics{nicfig5b.ps}
\caption{{\bf N.B. This is not the full resolution figure which is available
at astro.ic.ac.uk/Research/extragal/dla.html.}
Reduced frames showing the images of four of the targets
before and after psf subtraction. The pixel size is $0\farcs075$. The
targets are, from top to bottom, nos 1. CS\,73, 2. PC\,0056+0125,
3. PHL\,1222, 4. PKS\,0201+113. In each case the left image is the
pixel $S/N$ frame formed by dividing the data frame by the error frame,
and shows the full mosaic field $25\farcs0\times22\farcs2$. The
compass points N with the arm to the E. The length of the arm is
$3\arcsec$. The central image is a $\times 3$ zoom of the
$7\farcs5\times7\farcs5$ region centred on the target, after psf
subtraction. The vertical bar is of length $1\arcsec$. This shows the
$S/N$ frame for the small detection filter i.e. the $S/N$ within an
aperture of diameter 0.45\arcsec centred at that point. Peaks above
$S/N=6$ are compact candidates. To detect diffuse candidates a
corresponding frame for an aperture of diameter 0.9\arcsec was
constructed.  The right-hand frame identifies the candidates, listed in
Table 4. The symbol size scales with brightness. Filled symbols are
compact candidates, open symbols diffuse. Brighter objects are plotted
at the correct deconvolved orientation and ellipticity (Table 5),
while fainter sources are plotted as circles. The cross marks the
quasar position.}
\end{figure*}

\begin{figure*}
\vspace{20.8cm}
\includegraphics{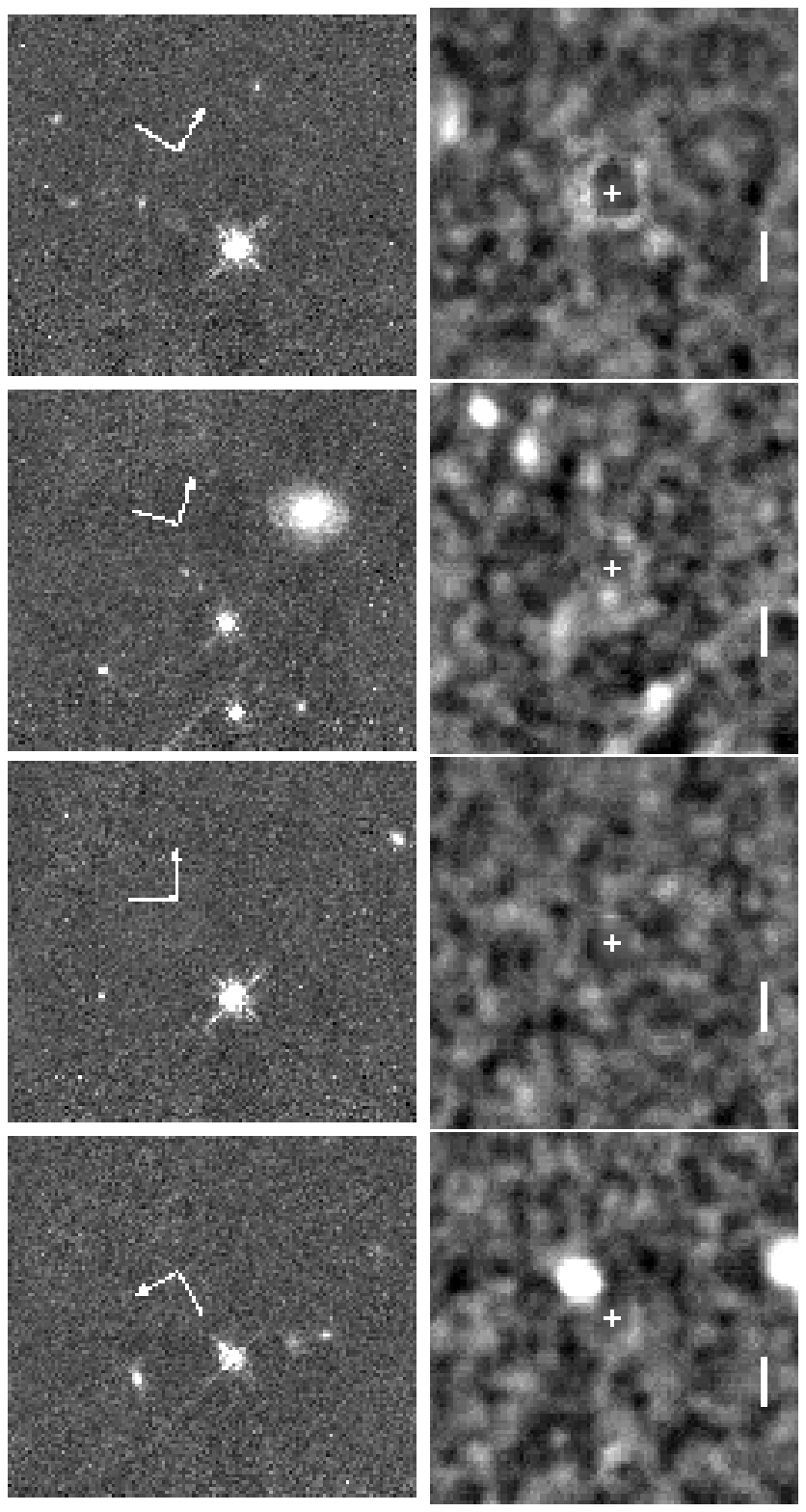}
\includegraphics{nicfig6b.ps}
\caption{{\bf N.B. This is not the full resolution figure which is available
at astro.ic.ac.uk/Research/extragal/dla.html.} Reduced frames showing the images of four of the targets
before and after psf subtraction. The pixel size is $0\farcs075$. The
targets are, from top to bottom, nos 5. 0216+0803, 6. PKS\,0458$-$02,
7. PKS\,0528$-$250, 8. H\,0841+1256. In each case the left image is
the pixel $S/N$ frame formed by dividing the data frame by the error
frame, and shows the full mosaic field
$25\farcs0\times22\farcs2$. The compass points N with the arm to the
E. The length of the arm is $3\arcsec$. The central image is a $\times
3$ zoom of the $7\farcs5\times7\farcs5$ region centred on the target,
after psf subtraction. The vertical bar is of length $1\arcsec$. This
shows the $S/N$ frame for the small detection filter i.e. the $S/N$ within
an aperture of diameter 0.45\arcsec centred at that point. Peaks above
$S/N=6$ are compact candidates. To detect diffuse candidates a
corresponding frame for an aperture of diameter 0.9\arcsec was
constructed. The right-hand frame identifies the candidates, listed in
Table 4. The symbol size scales with brightness. Filled symbols are
compact candidates, open symbols diffuse. Brighter objects are plotted
at the correct deconvolved orientation and ellipticity (Table 5),
while fainter sources are plotted as circles. The cross marks the
quasar position.}
\end{figure*}

\begin{figure*}
\vspace{20.8cm}
\includegraphics{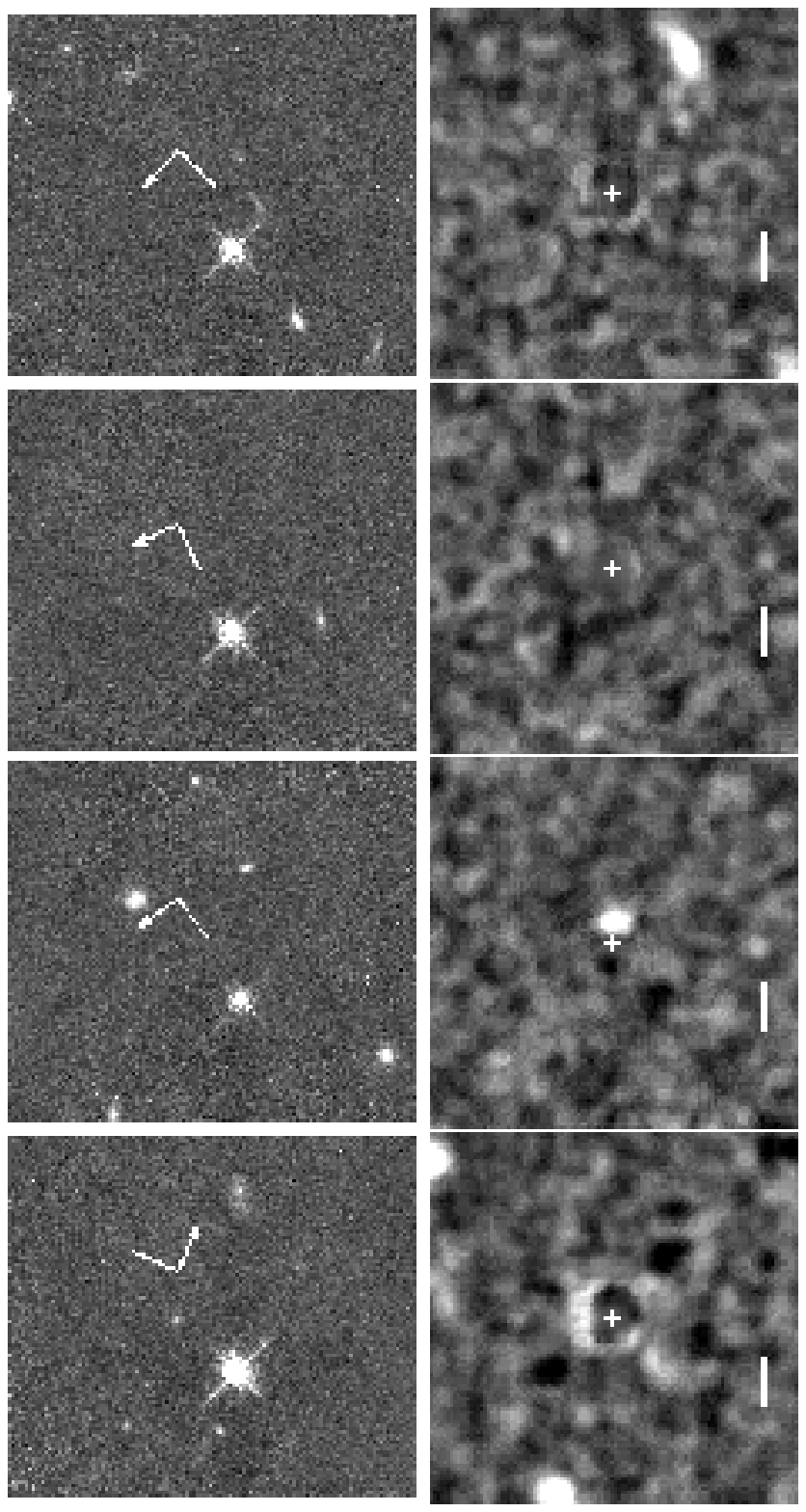}
\includegraphics{nicfig7b.ps}
\caption{{\bf N.B. This is not the full resolution figure which is available
at astro.ic.ac.uk/Research/extragal/dla.html.} Reduced frames showing the images of four of the targets
before and after psf subtraction. The pixel size is $0\farcs075$. The
targets are, from top to bottom, nos 10. B2\,1215+33,
11. Q\,1223+1753, 12. H\,1500$\#$13, 13. Q\,2116$-$358. In each case
the left image is the pixel $S/N$ frame formed by dividing the data
frame by the error frame, and shows the full mosaic field
$25\farcs0\times22\farcs2$. The compass points N with the arm to the
E. The length of the arm is $3\arcsec$. The central image is a $\times
3$ zoom of the $7\farcs5\times7\farcs5$ region centred on the target,
after psf subtraction. The vertical bar is of length $1\arcsec$. This
shows the $S/N$ frame for the small detection filter i.e. the $S/N$ within
an aperture of diameter 0.45 \arcsec centred at that point. Peaks
above $S/N=6$ are compact candidates. To detect diffuse candidates a
corresponding frame for an aperture of diameter 0.9\arcsec was
constructed. The right-hand frame identifies the candidates, listed in
Table 4. The symbol size scales with brightness. Filled symbols are
compact candidates, open symbols diffuse. Brighter objects are plotted
yat the correct deconvolved orientation and ellipticity (Table 5),
while fainter sources are plotted as circles. The cross marks the
quasar position.}
\end{figure*}

\begin{figure*}
\vspace{15.8cm}
\includegraphics{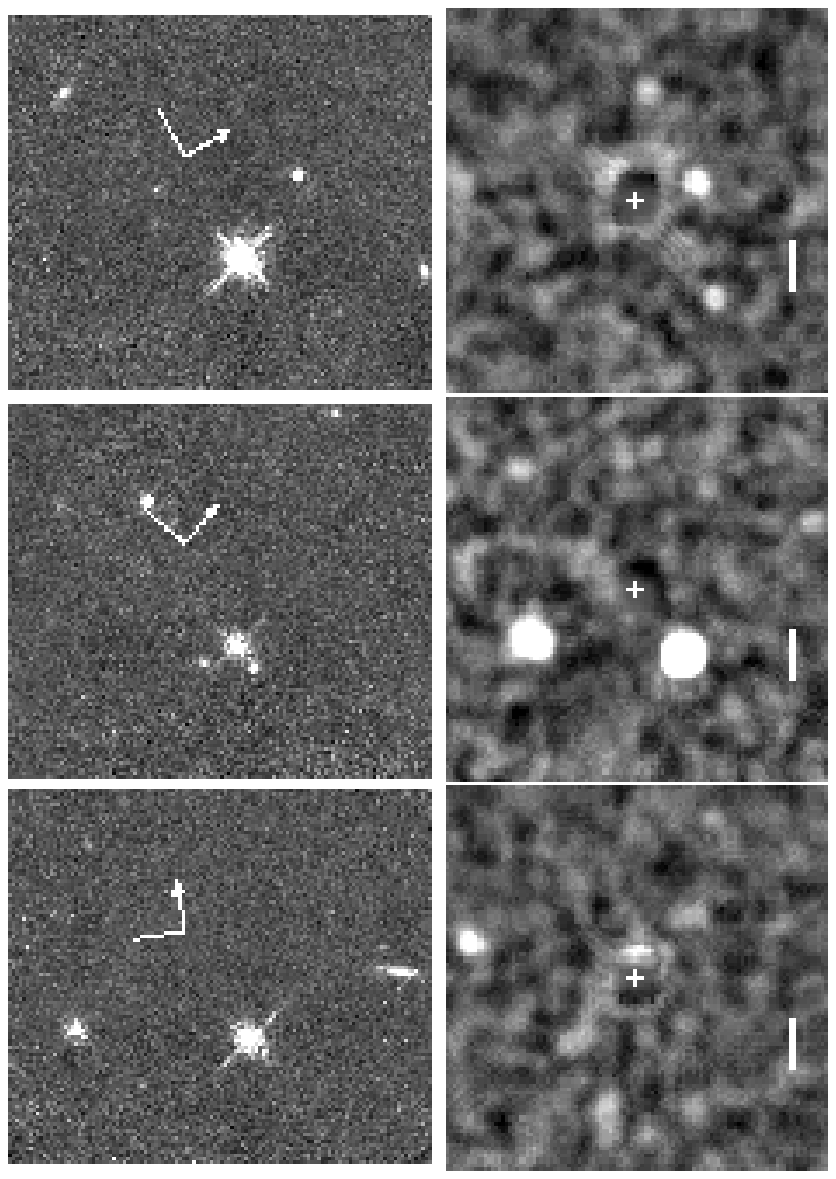}
\includegraphics{nicfig8b.ps}
\caption{{\bf N.B. This is not the full resolution figure which is available
at astro.ic.ac.uk/Research/extragal/dla.html.} Reduced frames showing the images of three of the targets
before and after psf subtraction. The pixel size is $0\farcs075$. The
targets are, from top to bottom, nos 14. Q\,2206$-$1958,
15. BR\,2212$-$1626, 16. 2233.9+1318. In each case the left image is
the pixel $S/N$ frame formed by dividing the data frame by the error
frame, and shows the full mosaic field
$25\farcs0\times22\farcs2$. The compass points N with the arm to the
E. The length of the arm is $3\arcsec$. The central image is a $\times
3$ zoom of the $7\farcs5\times7\farcs5$ region centred on the target,
after psf subtraction. The vertical bar is of length $1\arcsec$. This
shows the $S/N$ frame for the small detection filter i.e. the $S/N$ within
an aperture of diameter 0.45\arcsec centred at that point. Peaks above
$S/N=6$ are compact candidates. To detect diffuse candidates a
corresponding frame for an aperture of diameter 0.9\arcsec was
constructed. The right-hand frame identifies the candidates, listed in
Table 4. The symbol size scales with brightness. Filled symbols are
compact candidates, open symbols diffuse. Brighter objects are plotted
at the correct deconvolved orientation and ellipticity (Table 5),
while fainter sources are plotted as circles. The cross marks the
quasar position.}
\end{figure*}

\subsubsection{Results}

We now comment on the results for the individual fields in turn.

{\it 1. CS\,73, Fig. 5.} The spectrum of this source shows a LLS at
$z=1.8862$ and a DLA at $z=2.0713$ (Table 1), but no candidate
counterparts were found in this field.

{\it 2. PC\,0056+0125, Fig. 5.} There is a strong DLA line at
$z=2.7771$ in the spectrum of this quasar. Bearing in mind the small
impact parameters of the few confirmed DLA counterparts we consider
the nearest candidate, N-2-1C which lies at an angular separation of
2.32\arcsec, to be the most likely of the four candidate counterparts
listed in Table 4. The large ellipticity, 0.70, of the candidate
N-2-4D suggests that this is a late-type galaxy viewed at a high
inclination angle. The orientation of the image is only 10$^{\circ}$
from the line joining the quasar to the galaxy. Therefore a possible
alternative explanation of the DLA line is the presence of an extended
gaseous disk around this galaxy.

{\it 3. PHL\,1222, Fig. 5.} The redshift of the DLA $z=1.9342$ is
close to the redshift of the quasar $z=1.922$. The DLA absorbs much of
the quasar \lya emission line, and to measure the column density it
was necessary to model the unabsorbed \lya emission line profile (M\o
ller et al 1998). There are significant residuals visible after the
psf subtraction, which are approximately circularly symmetric, which we
suggest are most likely to be due to the quasar host galaxy. The DLA
absorber has been detected in \lya emission by Fynbo
et al. as an extended object 6 arcsec across, centred 0.9\arcsec E of
the quasar.  These authors suggest that the source of ionising photons
is the nearby quasar Q\,0151+0448B, rather than young stars. This
quasar lies at an angular separation of 3.32\arcsec, and is listed in
Table 4 as source N-3-2C. We do not find a source coincident with the
centre of the \lya emission, consistent with the photoionisation
explanation.

{\it 4. PKS\,0201+113, Fig. 5.} The spectrum of this source shows a
DLA absorber at $z=3.3875$. There is a relatively bright compact source
N-4-1C with ${\mathrm H_{AB}=23.36}$ only 1.01\arcsec from the quasar,
which is a good candidate for the counterpart. This object has $n=1.21$
i.e. the surface-brightness profile approximates an exponential.

{\it 5. 0216+0803, Fig. 6.} The spectrum of this source shows a
confirmed DLA absorber at $z=2.2930$. There is also a high-column
density system at $z=1.7688$ confirmed as such by Lu et al. (1996) on
the basis of the detection of the weak metal lines of Fe\,II
$\lambda$2260, Mn\,II $\lambda$2576, and Ni\,II $\lambda$1751. The
value for the HI column density however comes from the the \lya line
equivalent width measured by Lanzetta et al (1991) from a low
resolution spectrum, and is therefore uncertain. The nearest candidate
counterpart N-5-1C lies at an angular separation of 1.41\arcsec. As
noted in Table 4 this object lies on a diffraction spike, and
therefore should be considered less reliable than other sources,
especially given the relatively low $S/N$ of 6.3.

{\it 6. PKS\,0458$-$02, Fig. 6.} This quasar shows extended radio
emission. Analysis of the 21 cm line in absorption allowed Briggs et
al (1989) to place a lower limit on the size of the DLA absorber
($z=2.0395$) of $8h^{-1}$ kpc. The nearest candidate counterpart is
the faint ${\mathrm H_{AB}=24.60}$ diffuse object N-6-1D, at an
angular separation of only 0.86 arcsec. This candidate is the closest
to the quasar of all the candidates listed in Table 4, and very close
to our inner radius limit of 0.79 arcsec for the detection of diffuse
objects. The object does not stand out clearly in Fig. 6 (middle
column) as this is the detection frame for compact objects. We see no
reason to doubt the reality of this object. The source N-6-3C lies on
the diffraction spike of a nearby bright star, and the measured
magnitude will overestimate the brightness of this source.

{\it 7. PKS\,0528$-$250, Fig. 6.} There are two DLA absorbers in the
spectrum of this quasar at $z=2.1408$ and $z=2.8110$. The galaxy
counterpart of the higher redshift system was discovered by M\o ller
and Warren (1993) and has been extensively studied (Warren and M\o
ller 1996, M\o ller and Warren 1998). This source is detected in our
NICMOS image at $S/N=5.0$, and we have included the object in Table 4
even though it is fainter than the detection limit $S/N=6$.  The STIS
image of the source is presented in M\o ller et al (in preparation). The
$z=2.8110$ DLA system absorbs the quasar \lya emission line.  The
column density quoted for this absorber in Table 1 is taken from M\o
ller and Warren (1993) who attempted to account for the quasar \lya
emission in making the fit (in the same manner as with PHL\,1222).

{\it 8. H\,0841+1256, Fig. 6.} The redshift of H\,0841+1256 is
uncertain. The object was discovered by Hazard in a search of an
objective prism plate, and may be a BL Lac as no emission lines have
been detected. The redshift is estimated from the wavelength of the
onset of absorption in the \lya forest, but this cannot be measured
accurately as it coincides with the wavelength of the $z=2.4764$ DLA
absorber, which also absorbs any \lya emission that might be
present. There are two confirmed DLA absorbers in the spectrum of this
object, at $z=2.3745$ and $z=2.4764$.  A candidate DLA absorption line
near 3480\AA\, remains to be confirmed.

The source N-8-1C was first detected by Arag\'{o}n-Salamanca, Ellis,
and O'Brien (1996) in a K-band image obtained at the United Kingdom
Infrared Telescope (UKIRT). They measured ${\mathrm K=19.9}$ (Johnson)
for this source. Taking the total magnitude from Table 5 and
converting to the Johnson system, our measurement gives ${\mathrm
H=20.7}$, and a colour ${\mathrm H-K=0.8}$.

{\it 9. PC 0953+4749.} As stated earlier the NICMOS observations of
this source failed. The redshifts and column densities of the three
DLA absorbers were measured from a spectrum taken with the Keck I
telescope. The spectrum will be presented in Bunker et al. (in
preparation) together with WFPC2 observations of the field.

{\it 10. B2\,1215+33, Fig. 7.} There is a single DLA absorber at
$z=1.9990$ in the spectrum of this quasar. This field was also
observed by Arag\'{o}n-Salamanca et al (1996) who report the detection
of a candidate with ${\mathrm K=20.1}$ (Johnson) located $1.3\arcsec$
E of the quasar. This candidate therefore is 0.2 mag fainter in K than
N-8-1C. However N-8-1C has an aperture magnitude of ${\mathrm
H_{AB}=22.5}$ whereas we do not detect their fainter candidate above
our detection thresholds of ${\mathrm H_{AB}=25.7}$ and 24.9 for
compact and diffuse sources respectively. If the object had a similar
colour and surface brightness profile to N-8-1C we would have expected
to find a compact source with an aperture magnitude of ${\mathrm
H_{AB}=22.7}$. This implies that their candidate, if real, is
exceptionally red.

{\it 11. Q\,1223+1753, Fig. 7.} There is a single DLA absorber at
$z=2.4658$ in the spectrum of this quasar, but no candidate
counterparts have been detected. Several of the images of this field
suffer from high background noise (\S2.2.1).

{\it 12. H\,1500$\#$13, Fig. 7.} There is a DLA absorber at $z=3.1714$
seen in the spectrum of this source. After psf subtraction a strong
residual to the SW and close to the quasar remains, which is clearly
visible in Fig. 7. The centroid of the residuals lies inside our
detection radius limit, and the tangential orientation of the residual
leads us to believe that it is not due to a galaxy counterpart of the
absorber, but could possibly be emission from the quasar host
galaxy.

{\it 13. Q\,2116$-$358, Fig. 7.} The spectrum of this quasar shows a
LLS absorption line at $z=1.9966$. This is the second brightest quasar
in our list. The psf subtraction for this quasar is particularly poor,
leading to strong negative residuals for two of the diffraction
spikes. This is the only object for which such a pattern of residuals
is seen, suggesting that the observing configuration was different in
some way. Although we have been unable to identify a cause we note
that this quasar was the first to be observed in our campaign.  A ring
of excess emission is seen in Fig. 7 which could be due to the host
galaxy of the quasar. The candidate source closest to the quasar lies
on a diffraction spike and therefore should be considered less
reliable than other sources.

{\it 14. Q\,2206$-$1958, Fig. 8.} The nearest source to the quasar
N-14-1C, which lies at an offset of 1.13\arcsec, has been confirmed by
us as the counterpart to the DLA absorber at $z=1.9205$ (M\o ller et
al., in preparation). The source lies on a diffraction spike. The psf
subtraction for this source is comparatively poor, so the uncertainty on the
magnitude is larger than that indicated by the $S/N$ quoted in Table
4. The spectrum of this quasar shows a second DLA absorber at
$z=2.0762$.

{\it 15. BR\,2212$-$1626, Fig. 8.} There is a LLS absorber at
$z=3.6617$ in the spectrum of this quasar. There are two relatively
bright candidates in this field. With FORS on the VLT we have obtained
a spectrum of the brightest source N-15-2C, which lies at an angular
separation of 1.53\arcsec. This spectrum shows that the candidate
source is a quasar at the same redshift as BR\,2212$-$1626 (the
spectrum will be presented in a future paper describing spectroscopy
of our NICMOS and STIS candidates). While the hypothesis requires
further investigation at present we consider it unlikely that this is
a gravitational lens system, as we see no sign of a lensing galaxy in
our NICMOS image.

{\it 16. 2233.9+1318, Fig. 8.} The faint diffuse source N-16-1D with
${\mathrm H_{AB}=25.05}$ is the counterpart of the LLS at
$z=3.1501$. This object was first detected, and proposed as the
couterpart, by Steidel, Pettini, and Hamilton (1995), who measured an
offset from the quasar of $2.9\arcsec$, in agreement with our value
of $2.78\arcsec$. The object was discovered independently by
Djorgovski et al (1996), who confirmed the identification through the
spectroscopic detection of \lya emission. They quote a smaller offset
of $2.3\arcsec$. This object has also been detected in our STIS image
(M\o ller et al, in preparation). As with the quasar H1500$\#$13 after
psf subtraction there is residual emission close to the quasar, inside
our detection radius limit, which we do not believe is associated
with the absorber.

\subsubsection{Comparison with other work}

Kulkarni et al. (2000) observed the quasar LBQS 1210+1731 with the
NIC2 camera and F160W filter for a single orbit (2560 sec integration
time), and detected a candidate DLA absorber counterpart at an impact
parameter of $0.25\arcsec$, at $4.1\sigma$ significance. The object is
compact and would have an aperture magnitude of ${\mathrm
H_{AB}=21.87}$. This object is rather bright. Placing it in Fig. 4 it
can be seen that it is more than three magnitudes brighter than the
three spectroscopcially confirmed DLA galaxy counterparts
(${\mathrm H_{AB}=25.05, 25.11, 25.54}$), indicated by triangles.

As seen in Fig. 4, our average detection limit becomes rapidly
brighter at small impact parameters. Nevertheless their object lies
well above our $6\sigma$ detection limits. Our average limit for
compact objects at $0.3 \arcsec$ is ${\mathrm H_{AB}=23.87}$, with
range 22.49 to 25.33 (Table 3), so we would easily have detected a
similar object in any of our frames. Indeed we do find a few bright
candidates at small impact parameters. However we have chosen,
conservatively, not to list candidates inside $0.56\arcsec$ for the
reasons set out in section 2.4.

The accuracy of psf subtraction of the two methods appears to be
comparable. We compute an equivalent detection limit for their data of
${\mathrm H_{AB}=22.12}$, by scaling their measured noise at
$0.3\arcsec$ ($0.28\mu$Jy per pixel) to our longer integration
time. Their quasar has ${\mathrm H_{AB}=16.95\pm0.05}$ (total, P
Hewett, private communication), similar in brightness to our two
brightest targets, quasars 13 and 14. Our detection limits for these
quasars are 22.72 and 22.49.

\subsection{Summary}

In this paper we have described a search for the galaxy counterparts
of 23 high-redshift high-column density Ly$\alpha$ absorbers seen in
the spectra of 16 quasars, using the HST NICMOS instrument. Within a
box of side $7.5\arcsec$ centred on each quasar, over all the fields
we have found a total of 41 candidates, of which 3 have already been
confirmed spectroscopically as the counterparts. We provide detection
limits as a function of impact parameter for each field, and the use
of aperture magnitudes makes it very simple to compute whether a
hypothetical galaxy of specified luminosity profile and impact
parameter would have been detected in our survey.  Our
aperture-magnitude detection limits ${\mathrm H_{AB}\sim 25}$, the
small minimum impact parameter $0.56\arcsec$, and the sample size make
this the most sensitive search yet made for the galaxies producing
high-redshift DLA absorption lines. At the same time we are obtaining
very-deep optical images using STIS, which will provide an additional
list of candidates (M\o ller et al., in preparation). Except for
unusually red galaxies the STIS images will reach substantially deeper
than the NICMOS images, and therefore can be used to check the
reliability of the candidate list provided in Table 4. A discussion of
the conclusions that may be drawn from the imaging data, and the
limited spectroscopic follow-up so far completed, is deferred to the
STIS paper.

The next stage of this programme is a campaign of spectroscopy to
measure the redshifts of the candidates, to identify which are
counterparts of the absorbers. Since the galaxies are mostly faint,
and close to the quasar, redshift determination will be difficult. The
best hope will be to detect Ly$\alpha$ emission. Although this line is
readily extinguished by dust, and therefore may be weak, it is likely
to be the easiest to detect because at this wavelength the light from
the quasar is removed by the absorber itself.

\section*{Acknowledgments}

We are grateful to Beth Perriello (our STScI Program Coordinator) and
Al Schultz (our STScI Contact Scientist) for help during the design
and execution of the programme, and to Mark Dickinson for detailed
guidance on how best to reduce the data.


\begin{thebibliography}{}

\bibitem[1996]{ar96} Arag\'{o}n-Salamanca A., Ellis R. S., O'Brien
K. S., 1996, MNRAS 281, 945 
\bibitem[1989]{br89} Briggs F. H., Wolfe A. M., Liszt H. S., Davis
M. M., Turner K. L., 1989, ApJ 341, 650
\bibitem[1999]{bu99} Bunker A. J., Warren S. J., Clements D. L.,
Williger G. M., Hewett P. C., 1999, MNRAS 309, 875
\bibitem[1999]{ci99} Ciotti L., Bertin G., 1999, A\&A 352, 447
\bibitem[1996]{dj96} Djorgovski S. G., Pahre M. A., Bechtold J., Elston
   R., 1996, Nature 382, 234
\bibitem[1999]{fy99} Fynbo J. U., M\o ller P., Warren S. J., 1999,
MNRAS 305, 849 
\bibitem[1997]{ha97} Haehnelt M. G., Steinmetz M., Rauch M., 1998, ApJ
495, 647
\bibitem[1997]{ho97} Fruchter A. S., Hook R. N., 2000, PASP, submitted
\bibitem[2000]{ku00} Kulkarni V. P., Hill J. M., Schneider G., Weymann
R. J., Storrie-Lombardi L. J., Rieke M. J., Thompson R. I., Jannuzi
B. T., 2000, ApJ in press
\bibitem[1991]{la91} Lanzetta K. M., Wolfe A. M., Turnshek D. A., Lu
   L., Mc\,Mahon R. G., Hazard C., 1991, ApJS 77, 1
\bibitem[1997]{lb97} Le Brun V., Bergeron J., Boiss\'{e} P., Deharveng
J. M., 1997, A\&A 321, 733
\bibitem[1998]{le98} Ledoux C., Petitjean P., Bergeron J., Wampler
E. J., Srianand R., 1998, A\&A 337, 51
\bibitem[1997]{lu96} Lu L., Sargent W. L. W., Barlow T. A., Churchill
C. W., Vogt S. S., 1996, ApJS 107, 475
\bibitem[1994]{lu94} Lu L., Wolfe A. M., 1994, AJ 108, 44
\bibitem[1993]{lu93} Lu L., Wolfe A. M., Turnshek D. A., Lanzetta
K. M., 1993, ApJS 84, 1
\bibitem[1994]{mo94}M\o ller P., Jakobsen P., Perryman M. A. C., 1994,
A\&A 287, 719
\bibitem[1993]{mo93} M\o ller P., Warren S. J., 1993, A\&A 270, 43
\bibitem[1996]{mo98a} M\o ller P., Warren S. J., 1998, MNRAS 299, 661
\bibitem[1996]{mo98b} M\o ller P., Warren S. J., Fynbo J. U., 1998,
A\&A 330, 19
\bibitem[1993]{mo80} Morton D. C., Chen J., Wright A. E., Peterson
B. A., Jauncey D. L., 1980, MNRAS 193, 399
\bibitem[1999]{pe99} Pei Y. C., Fall S. M., Hauser M. G., 1999, ApJ
522, 604
\bibitem[1994]{pt94} Pettini M., Smith L. J., Hunstead R. W., King
D. L., 1994, ApJ 426, 79
\bibitem[1997]{pt97} Pettini M., Smith L. J., King D. L., Hunstead R. W.,
1997, ApJ 486, 665
\bibitem[1997]{pr97} Prochaska J. X., Wolfe A. M., 1997, ApJ 474, 140
\bibitem[1991]{sc91} Schneider D. P., Schmidt M, Gunn J. E., 1991, AJ,
101, 2004
\bibitem[1995]{st95} Steidel C. C., Pettini M., Hamilton D., 1995, AJ
110, 2519
\bibitem[1998]{ve98} V\'{e}ron-Cetty M.-P., V\'{e}ron P., 1998, ESO
Scientific Report No. 18
\bibitem[1996]{wa96} Warren S. J., M\o ller P., 1996, A\&A 311, 25
\bibitem[1996]{wa96a} Warren S. J., Hewett P. C., Lewis G. F., 
M\o ller P., Iovino A., Shaver P. A., 1996, MNRAS 278, 139
\bibitem[1993]{wh93} White R. L., Kinney A. L., Becker R. H., 1993,
ApJ 407, 456 
\bibitem[1987]{wo87} Wolfe A. M., Turnshek D. A., Smith H. E., Cohen
R. D., 1986, ApJS 61, 249
\bibitem[1995]{wo95} Wolfe A. M., Lanzetta K. M., Foltz C. B., Chaffee
F. H., 1995, ApJ 454, 698


\end{thebibliography}
\end{document}